\documentclass{aa}  

\usepackage{graphicx}
\usepackage{txfonts}
\usepackage{tikz}
\usepackage{hyperref}

\usepackage{subcaption}
\begin{document} 

\newcommand{\solar}{$_\odot$}
\newcommand{\tento}[1]{$10^{#1}$}
\newcommand{\timestento}[2]{$#1 \times 10^{#2}$}
\newcommand{\MAa}{$M_{\rm Aa}$}
\newcommand{\MAbone}{$M_{\rm Ab1}$}
\newcommand{\MAbtwo}{$M_{\rm Ab2}$}
\newcommand{\aAaAb}{$a_{\rm Aab}$}
\newcommand{\aAbonebtwo}{$a_{\rm Ab1Ab2}$}
\newcommand{\RAa}{$R_{\rm Aa}$}
\newcommand{\RAbone}{$R_{\rm Ab1}$}
\newcommand{\RAbtwo}{$R_{\rm Ab2}$}
\newcommand{\iAaAb}{$i_{\rm AaAb}$}
\newcommand{\iAbonebtwo}{$i_{\rm Ab1Ab2}$}
\newcommand{\fAab}{$\phi_{\rm Aab}$}
\newcommand{\fAbonebtwo}{$\phi_{\rm Ab1b2}$}
\newcommand{\lfrac}{$f_{\rm Aa}/f_{\rm Total}$}
\newcommand{\lfracBa}{$f_{\rm Ba}/f_{\rm Total}$}
\newcommand{\lfracBb}{$f_{\rm Bb}/f_{\rm Total}$}

\newcommand{\kepler}{\textit{Kepler}}
\newcommand{\starshadow}{\texttt{STAR SHADOW}}
\newcommand{\pofour}{\texttt{period04}}
\newcommand\kms{\ifmmode{\rm km\thinspace s^{-1}}\else km\thinspace s$^{-1}$\fi}
\newcommand{\gssp}{\texttt{GSSP}}

   \title{KIC~4150611: A quadruply eclipsing heptuple star system with a g-mode period-spacing pattern}

   \subtitle{Eclipse modelling of the triple and spectroscopic analysis}

   \author{
   Alex Kemp\inst{1}\thanks{
    \email{alex.kemp@kuleuven.be}}
    \and
  Andrew Tkachenko\inst{1}
    \and
    Guillermo Torres\inst{2}
    \and
    Kre\v{s}imir Pavlovski\inst{3}
    \and
    Luc IJspeert\inst{1}
    \and
    Nadya Serebriakova\inst{1}
    \and
    Kyle Conroy\inst{4}
    \and
    Timothy Van Reeth\inst{1}
    \and
    David Latham\inst{2}
    \and
    Andrej Pr\v{s}a\inst{5}
    \and
    Conny Aerts\inst{1}
  }

   \institute{Institute for Astronomy (IvS), KU Leuven, Celestijnenlaan 200D, 3001, Leuven, Belgium
        \and
             Center for Astrophysics $\vert$ Harvard \& Smithsonian, 60 Garden St, Cambridge, MA 02138, USA
         \and
             Department of Physics, Faculty of Science, University of Zagreb, 10 000 Zagreb, Croatia
        \and
             Space Telescope Science Institute, 3700 San Martin Dr, Baltimore, MD 21218, USA
        \and
            Villanova University, Department of Astrophysics and Planetary Sciences, 800 Lancaster Ave., Villanova, PA 19085, USA
             }

   \date{}

 
  \abstract
   {KIC~4150611 is a high-order multiple composed of a triple system composed of the F1V primary (Aa), which is eclipsed on a 94.2d period by a tight 1.52d binary composed of two dim K/M dwarfs (Ab1, Ab2), which also eclipse each other; an 8.65d eccentric, eclipsing binary composed of two G stars (Ba, Bb); and another faint eclipsing binary composed of two stars of unknown spectral type (Ca and Cb). In addition to its many eclipses, the system is an SB3 spectroscopic multiple (Aa, Ba, and Bb) and the primary (Aa) is a hybrid pulsator, exhibiting high amplitude pressure and gravity modes. In aggregate, this richness in physics offers an excellent opportunity to obtain a precise physical characterisation for some of the stars in this system.}
   {In this work, we aim to characterise the F1V primary by modelling its complex eclipse geometry and disentangled stellar spectra, in preparation for a follow-up work focusing on its pulsations.}
   {We employ a novel photometric analysis of the complicated eclipse geometry of Aa to obtain orbital and stellar properties of the triple. We acquired 51 \textit{TRES} spectra at the Fred L. Whipple Observatory, calculating radial velocities and orbital elements of Aa (SB1) and the B binary (SB2). These spectra and radial velocities are used to perform spectral disentangling for Aa, Ba, and Bb. Spectral modelling is applied to the disentangled spectrum of Aa to obtain atmospheric properties.}
   {
    From our eclipse modelling we obtain precise stellar properties of the triple, including the mass ratios ($M_{\rm Aa}/(M_{\rm Ab1}+M_{\rm Ab2})=3.61\pm0.01$, $M_{\rm Ab1}/M_{\rm Ab2} = 1.113\pm0.001$), separation ratio ($a_{\rm Aab}/a_{\rm Ab1Ab2} = 21.81\pm 0.01$), orbital periods ($P_{\rm Aab} = 94.29486\pm0.00008$d, $P_{\rm Ab1Ab2} = 1.522248\pm0.000001$d), and stellar radii (\RAa\ $ = 1.64\pm0.06$~R\solar, \RAbone\ $ = 0.42\pm0.01$~R\solar, \RAbtwo\ $ = 0.38\pm0.01$~R\solar).
    Radial velocity fitting and spectral disentangling arrive at orbital elements for Aa, Ba, and Bb in excellent agreement with each other and with previous results in the literature. Spectral modelling on the disentangled spectrum of Aa provides constraints on the effective temperature ($T_{\rm eff} = 7280\pm70$~K), surface gravity (log$(g) = 4.14\pm 0.18$~dex), micro-turbulent velocity ($v_{\rm micro} = 3.61\pm0.19~\kms$), rotation velocity ($v \sin i = 127 \pm 4$~\kms), and metallicity ({[}M/H{]} $ = -0.23 \pm 0.06$), also in good agreement with previous spectral modelling. Particular attention is paid to the light fraction of Aa, which our spectroscopic analysis determines to be between 0.92 and 0.94, while our eclipse modelling prefers a lower light fraction of $0.84\pm0.03$, similar to the previous literature value of 0.85. However, the eclipse models are still able to obtain an excellent fit to the solution when constrained to light fractions between 0.92 and 0.96, while our spectroscopic analysis proves to be far more sensitive to the light fraction, leading us to conclude that the higher light fraction from spectroscopy is likely the correct solution.}
   {}


   \maketitle
%

\section{Introduction}

High-order multiples remain one of the more poorly understood aspects of stellar evolution \citep{toonen2020,hamers2021,toonen2022}. Difficulties presented by uncertain single and binary stellar physics -- including magnetic, tidal, and mass transfer effects -- are often compounded by the rarity of these systems and the need to consider dynamical effects. A better understanding of the few examples of these systems we have \citep{kirk2016,kostov2022,rappaport2022,rappaport2023,kostov2024} is needed to help constrain their complex models.

The rich physics in these systems can play to our advantage, aiding our understanding of stellar evolution. Although binary (or higher-order) systems can pose a challenge to many observational surveys and science objectives, often discarded as `contaminants', in some circumstances binarity can lead to precise constraints on fundamental stellar quantities. Visual (or astrometric) binaries can -- in principle -- allow for targeted observations of individual components and astrometric measurements of tangential orbital elements. Single- and double-lined spectroscopic binaries (commonly referred to as SB1 and SB2 binaries) can be used to obtain radial velocity (RV) curves and in some cases, `disentangled' spectra \citep{simon1994,hadrava1995} for at least some of the stars in the system. And finally, eclipsing binaries can give vital information on the inclination angles, radii, and -- combined with spectroscopy -- masses of the system's components. Access to these fundamental properties\footnote{Table 1 of \cite{southworth2020} presents a useful graphical summary of the information that can be obtained from astrometric, spectroscopic, and eclipsing binaries.} is immensely helpful to modelling efforts attempting to reproduce observables such as asteroseismic pulsations \citep[e.g.,][]{hambleton2013,keen2015,guo2019,sekaran2021}.

\begin{figure}
\centering
\begin{tikzpicture}
\draw (0,1) -- (0,2);
\filldraw[black] (0,1) circle (2pt) node[anchor=north]{Aa};
\filldraw[white] (0,0.1) circle (2pt) node[anchor=south,color=black]{F1V};
\draw (0,2) -- (1,2) -- (1,2.2)  node[left]{A} node[right]{94.2d} -- (1,3);

\draw (1,2) -- (2,2) -- (2,1.7) node[left]{Ab} node[right]{1.52d} -- (2,1.5);

\draw (2,1.5) -- (1.5,1.5) -- (1.5,1);
\filldraw[black] (1.5,1) circle (2pt) node[anchor=north]{Ab1};
\filldraw[white] (1.5,0.1) circle (2pt) node[anchor=south,color=black]{K/M?};

\draw (2,1.5) -- (2.5,1.5) --(2.5,1.0);
\filldraw[black] (2.5,1) circle (2pt) node[anchor=north]{Ab2};
\filldraw[white] (2.5,0.1) circle (2pt) node[anchor=south,color=black]{K/M?};

\draw (1,2) -- (1,3);

\draw[dashed] (1,3) -- (2.5,3) node[anchor=south]{1000yr?} -- (4,3);
\draw[dotted] (4,3) -- (5,3) node[anchor=south]{?}-- (6,3);

\draw (4,3) -- (4,2) -- (4,2.2) node[left]{B} node[right]{8.65d} -- (4,2) --(3.5,2) -- (3.5,1.5);
\draw (3.8,2) -- (3.5,2);

\draw (3.5,2) -- (3.5,1.5);
\filldraw[black] (3.5,1.5) circle (2pt) node[anchor=north]{Ba};
\filldraw[white] (3.5,0.6) circle (2pt) node[anchor=south,color=black]{G};

\draw (4,2) -- (4.5,2) -- (4.5,1.5);
\filldraw[black] (4.5,1.5) circle (2pt) node[anchor=north]{Bb};
\filldraw[white] (4.5,0.6) circle (2pt) node[anchor=south,color=black]{G};

\draw (6,3) -- (6,2.2) node[left]{C} node[right]{1.43d} -- (6,2) -- (5.5,2);

\draw (5.5,2) -- (5.5,1.5);
\filldraw[black] (5.5,1.5) circle (2pt) node[anchor=north]{Ca};
\filldraw[white] (5.5,0.6) circle (2pt) node[anchor=south,color=black]{?};

\draw (6,2) -- (6.5,2) -- (6.5,1.5);
\filldraw[black] (6.5,1.5) circle (2pt) node[anchor=north]{Cb};
\filldraw[white] (6.5,0.6) circle (2pt) node[anchor=south,color=black]{?};

\end{tikzpicture}
\caption{Reproduction of Figure 15 from \protect \cite{heliminiak2017}, summarising the system hierarchy and nomenclature of KIC~4150611. \cite{heliminiak2017} present astrometric evidence for an association between the A and B components, but otherwise the A, B, and C components can be considered dynamically independent.}
\label{fig:hierarchy}
\end{figure}
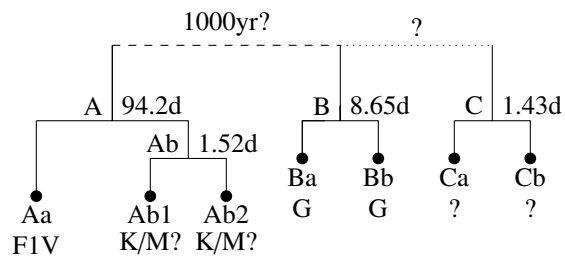

KIC~4150611 is a bright ($V\sim8$ mag) system with seven stars and four eclipsing components \citep{heliminiak2017}. Fig. \ref{fig:hierarchy} presents the system hierarchy and naming convention established by \cite{heliminiak2017}, who present the most recent and complete analysis of the system to date. The system is comprised of at least seven stars, although it should be noted that only the five stars forming the A+B system have direct evidence for dynamical association; the full A+B+C system can only be considered a candidate heptuple.

Aa, the primary, is a rapidly rotating ($128\pm5$ \kms) F1V-type \citep{niemczura2015} hybrid pulsator featuring prominent pressure  and gravity mode (henceforth p- and g-mode) pulsations of $\delta$-Scuti and $\gamma$-Dor type, respectively. Significantly, Aa's g-mode pulsations are arrayed in a period-spacing pattern of moderate length first identified in \citep{li2020a,li2020b}. This presents a valuable source of information, as the details of g-mode period-spacing patterns are sensitive to interior stellar properties such as the near-core rotation rate \citep{vanreeth2016}, and can be used in conjunction with grids of stellar models to constrain stellar properties such as stellar masses, ages, and mixing profiles \citep[see, for example,][]{aerts2018,johnston2019,pedersen2021,mombarg2021,michielsen2023}. The interested reader can refer to \cite{aerts2021} for a general review and introduction to asteroseismology, and \cite{guo2021,aerts2023} for reviews more focused around binary asteroseismology. Analysis of Aa's $\delta$-Scuti pulsations is performed in \cite{shibahashi2012,balona2014}.

Aa is in a 94.2d circular orbit with a tight 1.52d circular binary composed of two dim (negligible light contribution) dwarf\footnote{We denote these stars as M or K dwarfs in Fig. \ref{fig:hierarchy}, but note that this is an inference, not a spectroscopic result.} stars, Ab1 and Ab2. Ab1 and Ab2 eclipse both each other and Aa, sometimes simultaneously. It is the characterisation of this triple, particularly the pulsating primary Aa, that is the primary focus of this work. The other components of the system are the two G stars Ba and Bb, and two more stars of unknown spectral type Ca and Cb. \cite{heliminiak2017} present evidence for the B binary being dynamically associated with the A binary (on roughly a thousand year orbital period); there is no evidence one way or another as to whether the C binary is dynamically associated with the system. Ba and Bb form an eccentric SB2 eclipsing binary on an 8.65d orbit, while the C binary forms a tight 1.43d circular orbit. Ba and Bb account for most of the non-Aa light in the system, while the C binary is dim. Under adaptive optics, the A and B components of the system are spatially resolved at roughly 1 arc second apart, and a third, fainter source that may be the C component is also visible \citep[see Figure 1 of ][]{heliminiak2017}.

Other high-order multiples with numerous eclipses are known, such as the sextuply eclipsing sexptuple TIC168789840 analysed by \cite{powell2021}, and there are now hundreds of candidate triple/quadruple systems identified from \emph{TESS} data \citep{kostov2022,kostov2024}. However, KIC~4150611 remains exceptional in its potential for precise characterisation of its physical properties. Not only is it one of only $\approx 10$ \kepler\ stars with multiple ephemerides \citep{kirk2016}, it is also bright enough that phase-resolved spectroscopy of Aa, Ba, and Bb can be acquired. This already sets the quality of data available for KIC~4150611 above that of most other photometric high-order multiples. On top of this, the primary Aa is a hybrid pulsator with a g-mode period-spacing pattern, greatly assisting mode identification and therefore asteroseismic modelling \citep{aerts2018,aerts2021}. These patterns are rare, especially in combination with eclipses. Over 600 \kepler\ F-type stars were identified with g-mode period-spacing patterns by \cite{li2020a}, but only 35 eclipsing binaries with g-mode period-spacing patterns were identified in \cite{li2020b}, including KIC~4150611. Taken all together, KIC~4150611's combination of eclipses, spectroscopic viability, and asteroseismic information is extremely serendipitous.
 
This work is the first in a short series that builds on previous analyses of this fascinating system \citep[e.g.,][]{shibahashi2012,balona2014,heliminiak2017}. We focus on the eclipse modelling of the A triple and the spectral disentanglement of Aa, providing hard constraints on fundamental system properties. These constraints lay the groundwork for the second paper, which is dedicated to the asteroseismic modelling of Aa using its g-mode period-spacing pattern. The third and final paper plans to revisit the properties of the Ab, B, and C binaries using the modern \texttt{PHOEBE2} \citep{prsa2016,horvat2018,jones2020,conroy2020} software package.

\section{Methodology}

\subsection{Eclipse modelling}

\subsubsection{Light curve extraction for the eclipse modelling}


Light curve extraction and detrending is performed via a custom reduction of the \kepler\ long-cadence target pixel files (TPFs) intended to improve the available signal for the binary while minimising contaminants and the chance of signal being introduced during the detrending process. Similar detrending processes have been employed successfully in \cite{vanreeth2022,vanreeth2023}. \texttt{Lightkurve}\footnote{\hyperlink{https://docs.lightkurve.org/}{https://docs.lightkurve.org/}} was used to get all full-sector TPFs from the Mikulski Archive for Space Telescopes (MAST)\footnote{\hyperlink{https://archive.stsci.edu/}{https://archive.stsci.edu/}}. The raw light curve for each sector was then formed by adding pixels around the original \kepler\ mask with fluxes of at least 100 e$^-$/s when considering the 99th percentile of the measured flux in each pixel. Contamination from other stellar sources was avoided by halting the pixel-search upon finding any pixels that cause a rise in flux when moving outwards from the center of the mask. We includes plots of the TPF data and our masks for each quarter in the supplementary material available online (10.5281/zenodo.11505268, \hyperlink{https://zenodo.org/records/11505268}{https://zenodo.org/records/11505268}).

Because of the prominent eclipses and pulsations in this system, a somewhat complicated detrending process was necessary to isolate the instrumental effects when detrending the light curve. The raw light curves were first converted to mmag and each quarter was normalised using the median flux. This light curve was then detrended by first removing the 94.2d eclipses and then fitting a model composed of 20 sinusoids modelling the dominant frequencies (accounting for most of the 8.65d eclipse signal) and a linear model attempting to model the instrumental trend in each quarter. A linear model was selected to minimise the chance of introducing oscillatory signal into the light curve. The resulting sinusoidal component mostly models the strong 8.65d eclipses. This sinusoidal component is then subtracted from the data, removing almost all signal from the 8.65d eclipses, and we then refit the 20 dominant sinusoids (now modelling the strongest pulsations) and a new linear detrending model. This linear model is then finally subtracted from the original normalised data, forming our detrended, normalised light curve. It should be emphasised that the very rough sinusoid model developed during the detrending process is not used in any way other than to improve the accuracy of the detrending model.

The fully detrended light curve is then screened for 5-$\sigma$ outliers in an automated way, and then again through manual inspection. Finally, the 94.2d eclipses are once again removed from the light curve, and the remainder is prewhitened using two different methods: manually using \pofour\footnote{\hyperlink{http://period04.net/}{http://period04.net/}}, and in an automated way using \starshadow\ \citep{ijspeert2024}. 

Using \pofour, we first extract orbital harmonics of the 8.65d, 1.52d, and 1.43d eclipses, before proceeding to prewhiten the rest of the frequencies. During this process, groups containing 100 to 200 frequencies are simultaneously fit and prewhitened using \pofour's non-linear optimisation algorithm, before moving on to the next group. As we first extract orbital harmonics and then prewhiten more thoroughly, additional low-amplitude orbital harmonics become significant during the prewhitening process; these are identified and moved to the appropriate harmonic lists after all prewhitening is complete. In this way, we identify 178 harmonics of the 8.65d period, 18 harmonics of the 1.52d period, and 19 harmonics of the 1.43d period, in addition to 1238 other frequencies. 

\starshadow\ is a code package designed to obtain orbital and physical parameters for eclipsing binaries, and includes an automated pre-whitening algorithm highly suitable for such systems. This prewhitening process has five stages for a binary with known orbital period (it is also capable of identifying the orbital period in most cases; see \cite{ijspeert2024} for further details):

\begin{enumerate}
    \item Frequency identification is carried out through an iterative prewhitening process based on amplitude-hinting. The algorithm stops searching for additional frequencies when the Bayes information criteria (BIC) fails to reduce by at least 2 upon prewhitening the next frequency (implying that removing the frequency did not reduce the amount of information in the light curve by a significant amount). This step includes tests of whether there is a statistical benefit of keeping closely spaced frequencies as two distinct frequencies (versus a single frequency).
    \item The identified frequencies, amplitudes, and phases are then simultaneously optimised in (non-overlapping) groups of 20.
    \item Orbital harmonic frequencies are identified and their frequencies are fixed to integer multiples of the orbital frequency.  If desired, the orbital period can be determined at the start of this step.
    \item Additional frequencies are searched for among the residuals of this coupled-harmonic model. This step also re-applies the test relating to the statistical benefit of keeping closely spaced frequencies as two distinct frequencies (versus a single frequency).
    \item A final optimisation is computed, including all coupled harmonics as a single group, with other frequencies optimised in groups of 20.
\end{enumerate}

The need to deal with multiple orbital periods complicates matters somewhat. We first do a complete run through steps 1-5, coupling the 8.65d harmonics. Then these harmonics are subtracted from the light curve, with the residuals and remaining frequencies passed through steps 3-5 once more searching for the 1.52d orbital harmonics, and then a final time searching for the 1.43d orbital harmonics. We make use of \cite{heliminiak2017}'s orbital periods for this, which are sufficiently precise to predict the orbital harmonics with high accuracy. Without these pre-defined orbital periods, \starshadow\ was able to identify the 8.65d period harmonics with high accuracy, but not the far fainter 1.52d and 1.43d period harmonics.

At the end of this process we are left with three lists of orbital harmonics (corresponding to the 8.65d, 1.52d, and 1.43d orbital periods) in addition to the remaining, non-harmonic frequencies. This process results in 160 8.65d harmonics, 15 1.52d harmonics, and 12 1.43d harmonics, along with 884 other frequencies. The lower number of harmonics and other frequencies compared to the \pofour\ prewhitening is in part due to the amplitude-hinting method. Low frequency red noise in the periodogram results in the BIC stopping criteria being triggered before it reaches the low amplitude, high frequency peaks exposed during pre-whitening.

\begin{figure*}
\centering
   \includegraphics[scale=0.6]{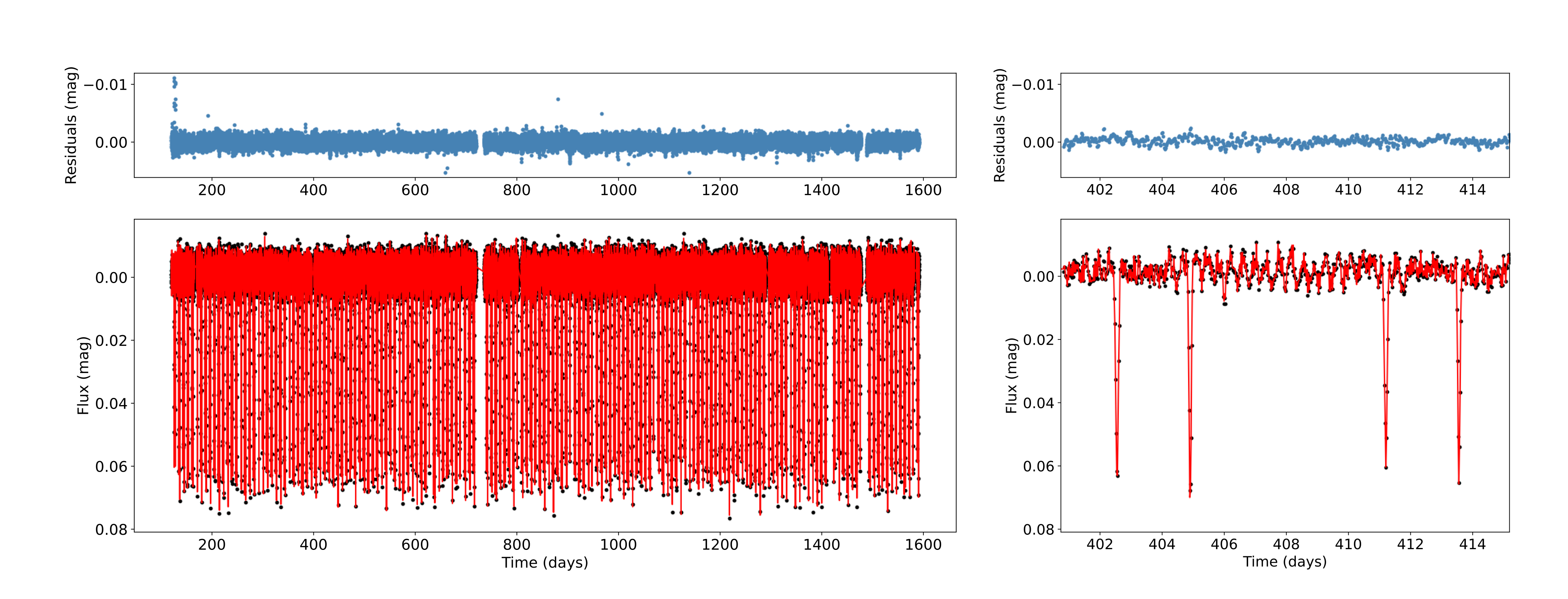}
       \caption{Normalised, detrended light curve (excepting 94.2d eclipses) and the sinusoid model formed from all frequencies extracted using \pofour.}
    \label{fig:pofourlightcurve}
\end{figure*}

These frequency lists will form the basis of our asteroseismic modelling in our next paper. In this work their importance lies in the fact that they each represent a sinusoidal model of the physical signal from the 8.65d, 1.52d, and 1.43d eclipses as well as the pulsations and much of the correlated noise present in the light curve. Our eclipse modelling models only the 94.2d eclipses of Aa. While the 1.52d Ab binary is part of our eclipse model, this is only considered in the context of the light blocked from Aa; the model does not include light from Ab1 and Ab2, and so the light variation stemming directly from their self-eclipses is not present in the model. Therefore, we subtract the \pofour\ model from the detrended 94.2d eclipse portions of the light curve to obtain a cleaned light curve that can be compared with our eclipse model. Subtracting the sinusoid model from \starshadow\ gives essentially the same residual light curve for the purposes of Aa eclipse modelling; the decision to work with the \pofour\ model can be considered arbitrary. The normalised, detrended light curve -- without the 94.2d eclipses -- and the \pofour\ sinusoid model is shown in Fig. \ref{fig:pofourlightcurve}.

\begin{figure}
   \centering
   \includegraphics[scale=0.4]{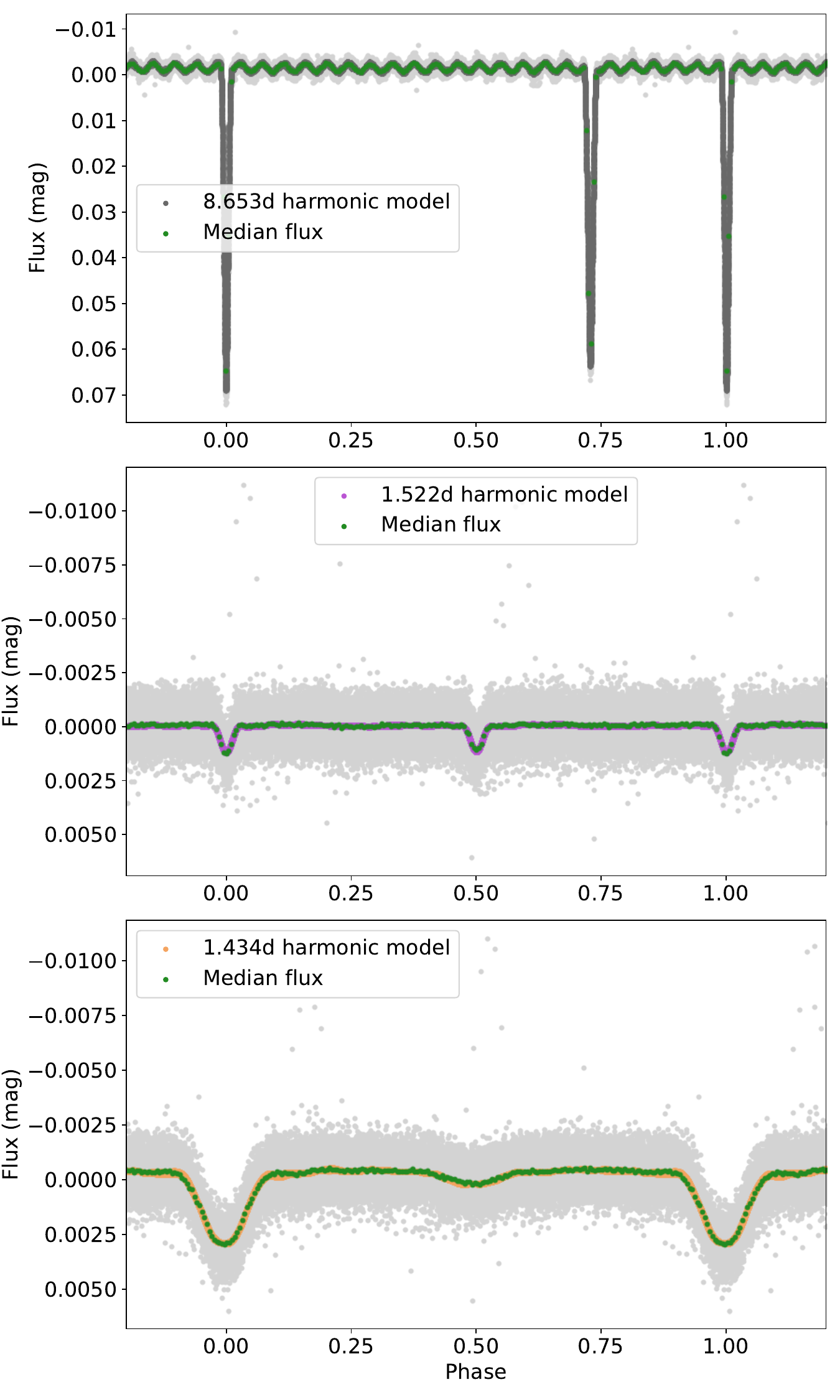}
   \caption{Phase-folded light curves for each of the eclipsing components and its sinusoid model. The light curve (grey data) for each is the residual between the normalised, detrended light-curve and every frequency except those forming the sinusoid model. See main text for discussion of the high frequency oscillatory signal present in the 8.65d model and the apparent discrepancy with \protect \cite{heliminiak2017} in the lower two panels.}
    \label{fig:pofourphasefold}
\end{figure}

We defer a more in-depth discussion of the impact of the two extraction methods on the asteroseismic analysis to our next paper, but for now we present the phase-folds of the various eclipsing components in Fig. \ref{fig:pofourphasefold} from the \pofour\ eclipse model.
The high frequency oscillatory signal present in the 8.65d \citep[see also Figure 2 of ][]{heliminiak2017} eclipses are an artefact of imprecise frequency extraction caused by a near-perfect coincidence between a g-mode pulsation and an 8.65d eclipse harmonic. This artefact does not affect our results in this work, as it is only present when considering \pofour's 8.65d eclipse model in isolation; it is not present when considering the full sinusoid model for the light curve. Making use of the coupled-harmonic methodology of \starshadow, both the correct harmonic and g-mode pulsation can be extracted, and the high frequency oscillatory behaviour disappears from the 8.65d eclipse model. 

A final comment on the phase folds is that when comparing with Figure 2 of \cite{heliminiak2017}, what we call the 1.52d eclipse is what that figure labels the 1.43d eclipse, and vice versa. This apparent conflict is due to  \cite{heliminiak2017} having mislabeled their figure; their caption is correct, in that their `S3' curve is indeed the middle panel, but as noted in Table 1 of \cite{heliminiak2017}, their S3 curve corresponds to the 1.52d orbital period. 
 
\subsubsection{Modelling the A triple}
\label{sec:meth:eclipsemodel}

The triple nature of KIC~4150611A dominates the behaviour of its eclipses. The combination of a large, bright F star on a long-period orbit with a tight dwarf-binary results in extremely complicated eclipse geometry, with the orbital motion of the eclipsing binary components Ab1 and Ab2 comparable to the orbital motion of the centre-of-mass of the Ab binary as a whole. This complexity precludes analytic eclipse analysis, and necessitates modelling the triple as a dynamical 3-body system.

We employ the n-body simulation package \texttt{REBOUND} \citep{rein2012,rein2015} to handle the dynamics of the problem. \texttt{REBOUND} has a number of integrators available, suitable for different scenarios; as all three components of the triple have non-negligible gravitational influence on each-other, we are forced to employ IAS15 \citep{everhart1985,rein2015}, the most accurate (and expensive) integrator.

The dynamics of the system taken care of, all that remains is the photometric model. For this, we are also able to take advantage of optimised `off-the-shelf' components, making use of parts of the exo-moon transit modelling code \texttt{Pandora} \citep{hippke2022}. Specifically, we use their flux calculation \texttt{occult.occult()}, which computes the flux of a bright body blocked by two dim bodies (here Ab1 and Ab2) which may also eclipse each other (this is relevant if it happens during an eclipse of the star (Aa), which it sometimes does for KIC~4150611). All three bodies are modelled as projections of spheres (circles) in the line of sight. The flux distribution of Aa is modelled using a quadratic limb darkening law, with limb darkening coefficients taken from \cite{claret2011} corresponding to a solar metallicity star with $T_{\rm eff}=$7400~K and log$(g)$=3.8 in the \kepler\ filter, consistent with \cite{niemczura2015}.

This simple photometry model allows us to consider the most relevant physics for this system (chiefly the multi-eclipsing dynamics), but does come with limitations. One restriction is that we can only consider spherical geometry. While Roche geometry is not relevant for this system, the rapid rotation of Aa will cause a small degree of oblateness, affecting the eclipse geometry. However, we expect the effect of this omission to be small, and we are able to obtain a satisfactory eclipse model without its inclusion. The treatment of Ab1 and Ab2 as dim bodies is not of concern for this system, nor, given the F1V spectral type, is the relatively simple quadratic limb darkening law.

\begin{figure}
   \centering
   \includegraphics[scale=0.40]{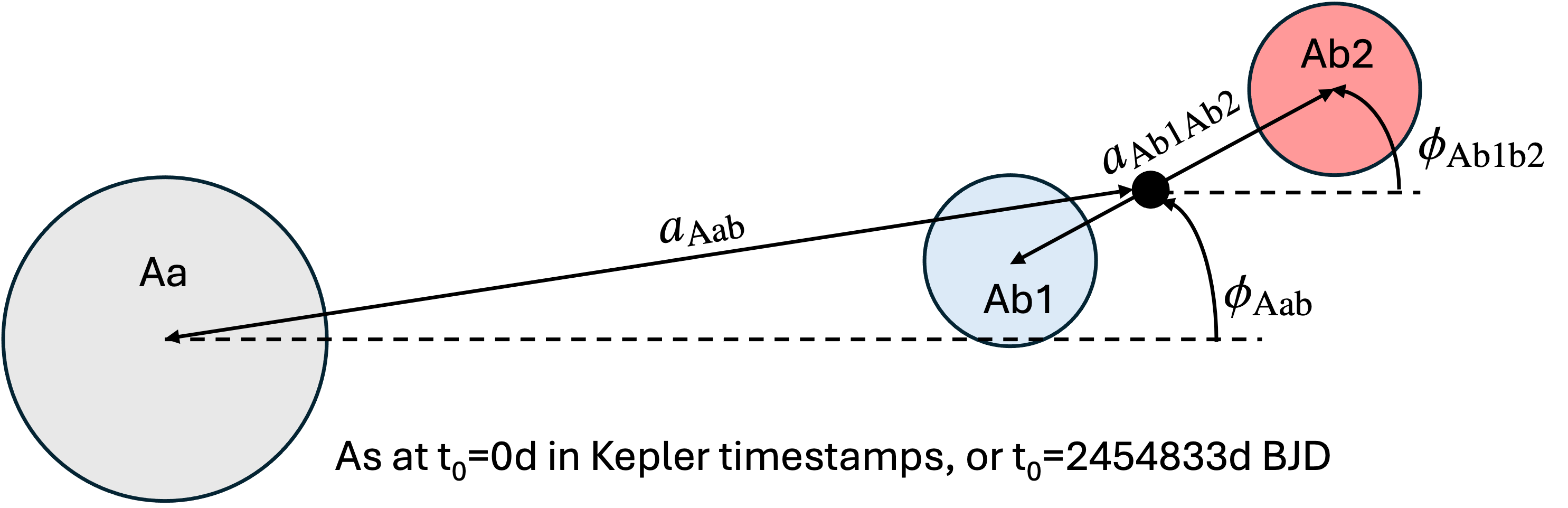}
   \caption{`Top-down' diagram illustrating how the separation and orbital phase terms in the n-body model are defined. The black dot represents the center of mass/rotation of the Ab binary. Both the A and Ab binaries orbit counter-clockwise from this view-point, and an observer would be at the bottom of the page.}
    \label{fig:triple_diagram}
\end{figure}

The n-body simulation provides a `fully physical' model, meaning that all modelled parameters directly correspond to physical quantities of the system. The modelled parameters are the masses (\MAa, \MAbone, \MAbtwo) and radii (\RAa, \RAbone, \RAbtwo) of each star, the inclination angles of each orbit (\iAaAb, \iAbonebtwo), the separations (\aAaAb, \aAbonebtwo) and phase terms (\fAab, \fAbonebtwo) at time $t_0=0$ in \kepler\ time stamps ($t_0=2454833$ BJD) of the A and Ab binaries, and finally the fraction of the system's total light output produced by Aa (\lfrac). The remainder of the light is assumed to be from components not part of the eclipse model (i.e. anything other than Aa, Ab1, or Ab2). Both binaries are initialised on perfectly circular orbits. Fig. \ref{fig:triple_diagram} illustrates the definition of the phase and separation terms.

Our ultimate goal is to provide constraints on the physical properties of this system.
As a first step, an initial estimation of reasonable bounds for the physical parameters was performed (see Table \ref{tab:eclipse_bounds}) based on the analysis by \cite{heliminiak2017}. For the special case of the light fraction, two different bounds were investigated to assess the sensitivity of our results to assumptions about this parameter.

The necessity of bounding the light fraction is driven by undesirable behaviour of the optimiser and Markov Chain Monte Carlo (MCMC) simulations when this parameter is left free from the start of the simulation. If \lfrac\ is left unconstrained and initialised with a sub-optimal fit that does not match the eclipse depths and/or positions well, these algorithms quickly reduce the light fraction to unrealistic levels -- 0.5-0.6 -- as they chase short-term gains in the objective function by reaching an average approximation of the eclipse depth, even as the models fail to match the eclipse shape. This biases subsequent steps attempting to tune the other parameters, which have more nuanced effects on the eclipse geometry. We deal with this by considering two light fraction bounds: 0.83-0.87 and 0.92-0.96,

\begin{table}[]
\centering
\begin{tabular}{lll}
Variable       & Range     & Units   \\
\hline
\MAa           & 1.2-1.8   & M\solar \\
\MAbone        & 0.35-0.55 & M\solar \\
\MAbtwo        & 0.25-0.45 & M\solar \\
\RAa           & 1.2-2.4   & R\solar \\
\RAbone        & 0.3-0.7   & R\solar \\
\RAbtwo        & 0.3-0.7   & R\solar \\
\aAaAb         & 106-118   & R\solar \\
\aAbonebtwo    & 4.8-6.0   & R\solar \\
\iAaAb         & 0-1       & $^{\circ}$ \\
\iAbonebtwo    & 0-2       & $^{\circ}$ \\
\fAab          & 4.0-4.3   & Radians \\
\fAbonebtwo    & 0-$2\pi$  & Radians \\
\lfrac & 0.83-0.87         & - \\
\lfrac & 0.92-0.96         & -

\end{tabular}
\caption{Bounds for the masses, radii, orbital separations, inclination angles (0 being an `edge-on' viewing angle), phase terms, and the light fraction for the eclipse modelling of KIC~4150611A. The repetition of the light fraction \lfrac, the fraction of total light contributed by Aa, is due to two different cases being considered for this parameter.}
\label{tab:eclipse_bounds}
\end{table}

\begin{figure*}
   \centering
   \includegraphics[scale=0.5]{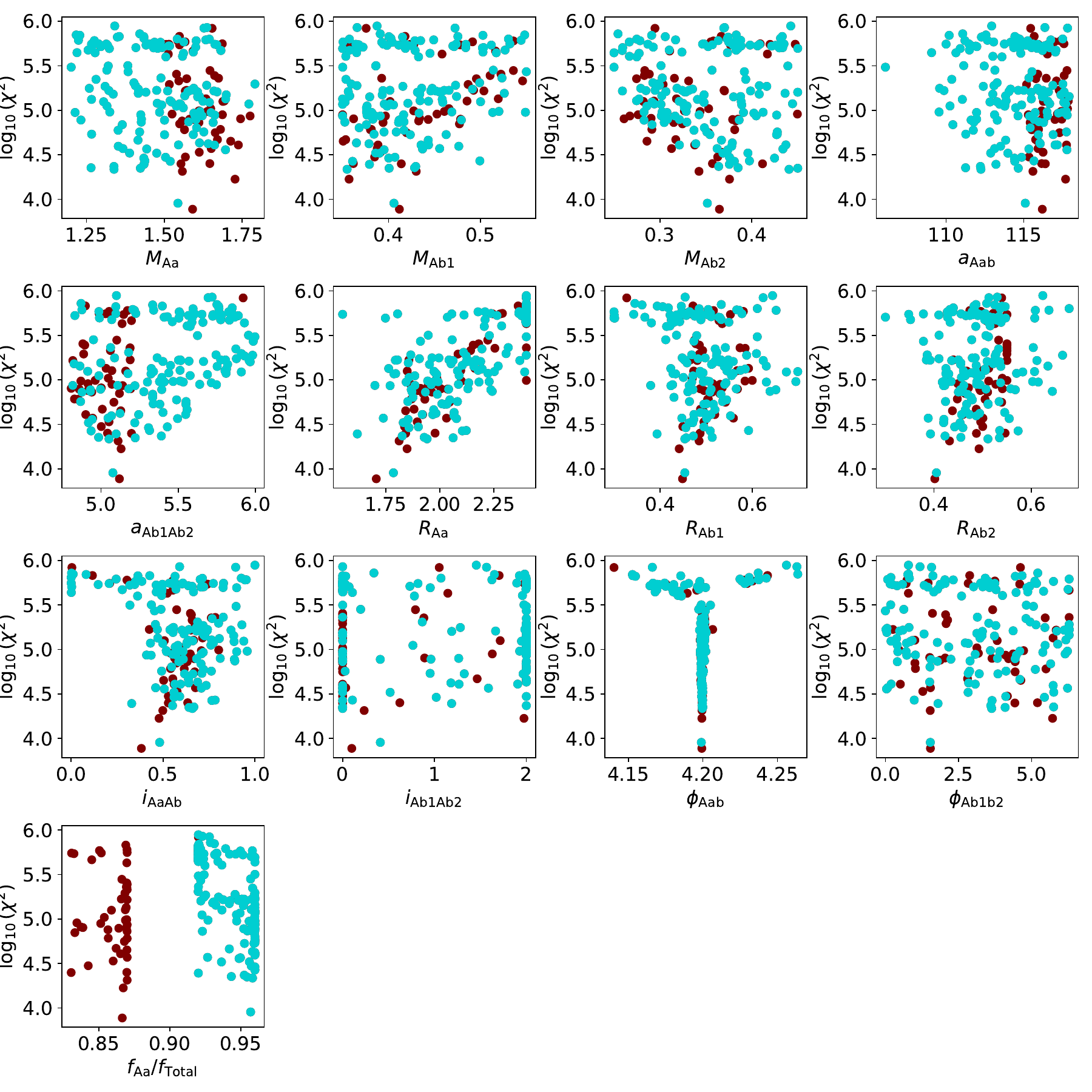}
   \caption{The results of 196 optimisation runs. We first compute 1000 steps of dual-annealing, the result of which is then used to initialise a Nealder-Mead optimisation stage. The best solution for the 0.83-0.87 light fraction (maroon) and 0.92-0.96 light fraction (blue) is then used to initialise a uniformly distributed `ball' of MCMC walkers, the results of which are presented in Section \ref{sec:res:eclipse}}
    \label{fig:solution_scatter}
\end{figure*}

This initial estimation was used to initialise a series of optimisation attempts to probe the parameter-space. Each optimisation attempt was computed by first running 1000 iterations of a stochastic dual-annealing algorithm resistant to local minima and then running a Nealder-Mead optimiser in an attempt to drill deeper into the best solution found by the dual-annealing algorithm\footnote{We make use of \hyperlink{Scypy}{https://scipy.org/}'s Nealder-Mead and dual annealing optimisation algorithms.}. This process was repeated 48 times for \lfrac\ between 0.83 and 0.87 light fraction, and 144 times for \lfrac\ between 0.92 and 0.96, which proved more prone to becoming stuck in local minima. The results of these optimisations are shown in Fig. \ref{fig:solution_scatter}. Fig. \ref{fig:solution_scatter} should not be used to infer properties of the landscape; a detailed discussion of the landscape can be found in Section \ref{sec:res:eclipse}.

The best model from this optimisation stage for each light fraction bound (0.83-0.87 and 0.92-0.96) was then used to initialise a MCMC simulation using \texttt{emcee} \citep{dfm2013}. These simulations serve both to drill deeper into the landscape and explore degeneracies and correlations between modelled parameters. Each MCMC simulation is initialised with 96 walkers stochastically distributed in a tight, uniform, 13-dimensional ball around the best result of the optimisation stage. The same parameter bounds described in Table \ref{tab:eclipse_bounds} were also enforced during these simulations. 

Each of these simulations was run for 350,000 steps, although the burn-in time was eventually estimated to be only a few tens of thousands of steps; the simulations were run so long to ensure that the degeneracies between the masses and separations were truly degeneracies, as opposed to simply being a collection of correlated parameters that had not converged. The results in this work present distributions computed using the last 50,000 steps of these simulations (i.e. we discard  300,000 steps as burn-in). Trace plots for the modelled parameters and the merit function for this dataset can be found in the online supplementary material.

To further explore the effect of our bounds on the light fraction, a third MCMC simulation was also run, initialised using the converged state of the 0.83-0.87 MCMC simulation. In this simulation, we allow the light fraction to become a free parameter. This simulation we evolve for 200,000 steps; results shown use the last 50,000 steps.

\subsection{Spectroscopic observations and radial velocities}
\label{sec:meth:specobs_rvs}

We gathered a total of 51 spectra of KIC 4150611 between July 2011 and July 2013, using the Tillinghast Reflector Echelle Spectrograph \citep[\textit{TRES};][]{szentgyorgyi2007} mounted on the 1.5m Tillinghast reflector at the Fred L.\ Whipple Observatory (Mount Hopkins, Arizona, USA). The spectra cover 51 orders between 3800 and 9100~\AA, at a resolving power of $R \approx 44,000$. The signal-to-noise ratios of the spectra range from about 70 to 340 per resolution element of 6.8~\kms.

All spectra are triple-lined, showing the lines of the F-type star (Aa) as well as those of both G-type stars (Ba, Bb). Radial velocities were derived with \texttt{TRICOR} \citep{zucker1995}, which is a three-dimensional cross-correlation algorithm that uses a different template for each component. The templates were taken from a large library of synthetic spectra based on Kurucz model atmospheres \citep[see][]{nordstroem1994,latham2002}, but with a line list tuned to better match real stars. The templates cover a 300~\AA\ region centered on the \ion{Mg}{1}b triplet around 5187~\AA. Radial velocities were obtained using the spectral order centered on that region, which contains most of the velocity information. For the bright F1V primary, we chose a template with a temperature of 7250~K, $\log g = 4.0$, and $v \sin i = 140~\kms$. For the two nearly identical G stars, we chose 5750~K, 4.5, and 8~\kms, respectively, which we found gave the best results. The heliocentric velocities we obtained are listed in Table \ref{tab:rvs_torres}, and are also available in machine-readable format in the online supplementary material.

\subsection{Spectral disentangling}

Spectral disentangling (SPD) is a method for reconstructing the individual stellar spectra from a time series of complex spectra of a binary/multiple system, allowing spectral modelling of the individual components of the binary. The method has the added advantage that disentangled spectra benefit from improved signal-to-noise, as this scales with the square root of the number of observations taken. In SPD it is assumed that the complex spectrum of a binary/multiple system is a linear combination of the individual components' spectra, each shifted by its radial velocity in the course of the orbit and diluted by the various components' fractional light contribution. This assumption generally holds except during the phases in which the line profiles might be distorted (e.g.~in the ingress or egress of the eclipses or blended with the interstellar lines and bands). The method was first formulated in wavelength (velocity) space by \cite{simon1994} and subsequently realised in Fourier space by \cite{hadrava1995}.

In the spectra of KIC~4150611, the dominant contribution is from the F-type star Aa, with traces of the spectra of the two G-type stars Ba and Bb also apparent. Aa is in a long-period triple system where the secondary components (Ab1 and Ab2) are not visible. The triple system A and binary system B are not in a hierarchical configuration, and therefore pose difficulty for SPD. There are two options: 

\begin{enumerate}
    \item Perform Fourier disentangling in two steps, first solving the A system as an SB1, and then, in the second step, use the residuals to generate the solution for the SB2 system of the two G-type stars.
    \item First measure the RVs for all three visible components (Aa, Ba, and Bb) and then perform SPD in velocity-space with the fixed RVs.
\end{enumerate}

We opted for the second option, directly separating the individual components' spectra with SPD following \cite{simon1994}. The first option was later used as a sanity check, confirming our results. The measurements of the RVs for the three visible stars are described in Section \ref{sec:meth:specobs_rvs} and provided in Table \ref{tab:rvs_torres}.

We treated the light fraction as unknown in our calculations; hence, SPD is performed in pure separation mode. In other words, SPD calculations realise individual spectra diluted in a common continuum of the visible stars. However, information on the fractional light contributions of each visible component is preserved, and can be extracted in the optimal fitting of disentangled (separated) spectra \citep{tamajo2011,tkachenko2015}. It has been found that a precision of $\sim 1$ percent or better could be achieved in the determination of the fractional light contribution from the optimal fitting of disentangled spectra \citep{pavlovski2009b,pavlovski2018,pavlovski2022,pavlovski2023}.
 
SPD is performed in wavelength space with the \texttt{CRES} code \citep{ilijic2004CRES}. From the set of 51 high-resolution \textit{TRES} spectra for KIC~4150611 we used all but spectrum 24, which was discarded for having insufficiently high S/N. SPD is performed for the spectral range from 4000 to 5800 {\AA}. This is, by necessity, done in short spectral segments to make the calculations possible; dealing with three components and high-resolution spectra is demanding on both memory and CPU time. As the observed spectra are of similar S/N we did not assign weights for individual exposures.
 
As a validation test for the disentangled spectra of the visible components, we also performed Fourier disentangling in two steps (option one) using the \texttt{FDBinary} code \citep{ilijic2004}. We found that the disentangled spectra as well as the SB1 (Aa) and SB2 (Ba and Bb) orbital parameters are in excellent agreement between the two disentangling methods.

\section{Results and Discussion}

\subsection{Eclipse modelling}

Here we present the results of our eclipse modelling. However, before proceeding, a comment on the light fraction of Aa is necessary.

As discussed in Section \ref{sec:meth:eclipsemodel}, we computed three MCMC simulations for different light fraction treatments. Overall, we find that based on our eclipse modelling, there is a small statistical benefit to a lower light fraction, but this conclusion is tenuous, particularly in the face of our spectral modelling on the disentangled spectra of Aa. Based on our spectroscopic analysis (Section \ref{sec:res:spec}), which appears to be far more sensitive to the light fraction, we consider a light fraction between 0.92 and 0.96 to be the most plausible. In this section, we will consider all three light-fraction cases in an attempt to assess the effect of the light fraction on the uncertainty of our results.

\subsubsection{Eclipse geometry}

\label{sec:res:eclipse}

\begin{figure*}
   \centering
   \includegraphics[scale=0.33]{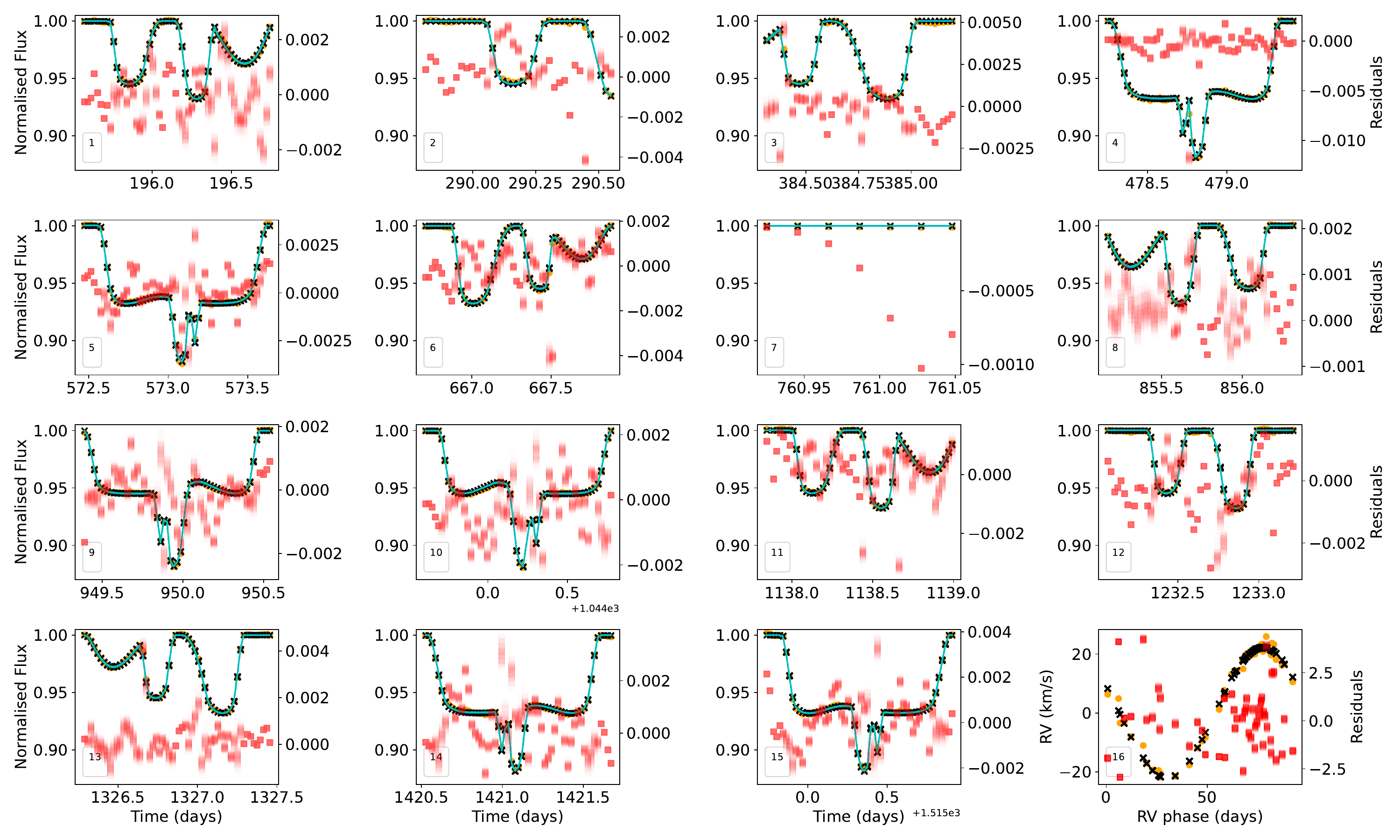}
   \caption{100 randomly selected models from the converged stage of our MCMC simulation for 0.92-0.96 \lfrac. Orange is long-cadence \kepler\ data, black/cyan is the model predictions, and red are the residuals at each timestamp. The primary y-axes are the normalised flux, with the secondary being the residuals. Rather than an eclipse, the bottom right panel shows the radial velocity curve in phase space.}
   \label{fig:esc9296}
\end{figure*}

Fig. \ref{fig:esc9296} shows 100 randomly selected models from the last 50,000 steps of our 0.92-0.96 \lfrac\ MCMC simulation overlaid with the relevant eclipsing portions of the light curve; to the naked eye these different models (cyan curves/black data) are nearly indistinguishable. The secondary y-axis shows the residuals (red, transparent squares) to give a better idea of the level of variation between the models.

Rather than an eclipse, panel 16 shows the simulated and observed radial velocity (RV) profile for Aa. The radial velocity profile for Aa was included as part of the objective function in both the optimisation and the MCMC stages of parameter exploration as an unweighted term (in \kms). However, even without this inclusion the fit to the radial velocity curve from the models is satisfactory given the error margins on the observations. Limited experimentation with increasing the weight (up to a factor of 100) given to the radial velocity measurements had a negligible effect on the overall fit both to the radial velocity curve and the eclipses. It is unsurprising that the radial velocity curve of Aa provided no significant additional constraining power; it is difficult to envision a scenario where a model could successfully reproduce the eclipses but not the radial velocity curve.

The resulting fit to both the eclipses and radial velocity curve is excellent. The larger residuals in the radial velocity panel can be entirely attributed to the naturally larger scatter in the radial velocity measurements compared to the cleaned eclipse light curves. The equivalent figure for 0.83-0.87 light fraction can be found online, and to the eye the fit is just as good (as expected given the nearly identical value of the converged merit function (630 vs 625) for both cases).

We now turn our discussion to the details of each eclipse. In the supplementary online material, we have included a small collection of animations (gifs) to illustrate the underlying causes of the different eclipse geometries.

Three different eclipse geometries can be identified; which occurs depends on the orbital phase of the Ab binary at the onset of the eclipse.

\begin{enumerate}
    \item \textbf{The triple eclipse.} There are two variants of this eclipse geometry. The first variant is when one of the stars in the Ab binary completely transits Aa, followed shortly by its Ab companion. The two Ab stars then eclipse, with the original transiting star now experiencing a brief period of retrograde motion which causes it to eclipse -- usually only partially -- Aa a second time. Panel 1 of Fig. \ref{fig:esc9296} is an example of this geometry. The second variant is when the  first star in the Ab binary -- usually only partially -- eclipses Aa just before its period of retrograde motion begins (ending the eclipse), followed by it eclipsing its Ab companion (off the Aa star) and its Ab companion then transitting Aa, followed shortly by the transit of the originally occulting Ab star. This looks like the first variant in reverse; panel 13 (lower left) of Fig. \ref{fig:esc9296} is an example of this geometry.
    
    \item \textbf{The double eclipse.} This is the simplest geometry, but actually appears to be quite rare for this orbital configuration. Here the first star in the Ab binary eclipses Aa, followed by the second star. The retrograde motion associated with the Ab eclipse causes neither star to eclipse Aa a second time (or at most only barely occult Aa, producing negligible flux variation). Panel 12 of Fig. \ref{fig:esc9296} is an example of this geometry.
    
    \item \textbf{The super eclipse.} This is the most striking geometry. The first Ab star begins its transit, but its retrograde motion leading up to its eclipse extends the duration of this transit significantly. While this star is eclipsing Aa, the second star in the Ab binary begins its transit, approximately doubling the light blocked from Aa. This geometry always results in the Ab binary eclipsing itself while eclipsing Aa, leading to a jagged spike in flux during the super eclipse caused by the reduction in covered area of Aa during the eclipse of Ab1 and Ab2. The prograde velocity of the second transiting star in the Ab binary is always at its maximum in this scenario, resulting in the short duration of its transit of Aa, just as the retrograde motion of the first transiting star extends its transit time. Panel 4 of Fig. \ref{fig:esc9296} is an example of this geometry. It could also be argued that there are two variants of this geometry, based on the relative position of the spike during the super eclipse. Panel 9, for example, shows the spike occurring earlier (to the left) while panel 10 shows the spike occurring later (to the right). The position of the spike is determined by where on the surface of Aa the first transiting star is when the Ab eclipse occurs; an early spike means that the Ab eclipse is occurring on the left half of Aa (with most of the duration of the second Ab star's transit still to come), while a later spike means that the Ab eclipse is occurring on the right half of Aa (with most of the duration of the second Ab star's transit already having occurred).
\end{enumerate}

As a test of the model, we also compared our eclipse models predictions against \textit{TESS} and short-cadence \kepler\ photometry. These comparisons are shown in Figs. \ref{fig:esc9296tess} and \ref{fig:esc9296sc} for the 0.92-0.96 light fraction case, with the other light fractions included in the supplementary online material. The agreement with both the TESS data and the short-cadence Kepler data is excellent, including the flux spike caused by the eclipse of the Ab binary during the super-eclipses. This strongly implies that formal inclusion of this data in the MCMC process -- which would significantly drive up the computational cost -- would have negligible effect on the parameter estimation.

\subsubsection{Parameter estimation}

Here we present and discuss the results of our parameter estimation. We will present the results of all three of our MCMC simulations, as it informs the discussion on the relevant uncertainties of the various parameters.

As discussed in Section \ref{sec:meth:eclipsemodel}, we rely upon a fully physical model. This does not mean we are naturally guaranteed constraints on all physical parameters. Without radial velocities for the Ab binary, absolute masses and orbital separations cannot be obtained. Instead of converging, these parameters slowly wander around the parameter-space in a correlated ball. The values of these parameters cannot be considered meaningful.

\begin{figure*}
   \centering
   \includegraphics[scale=0.25]{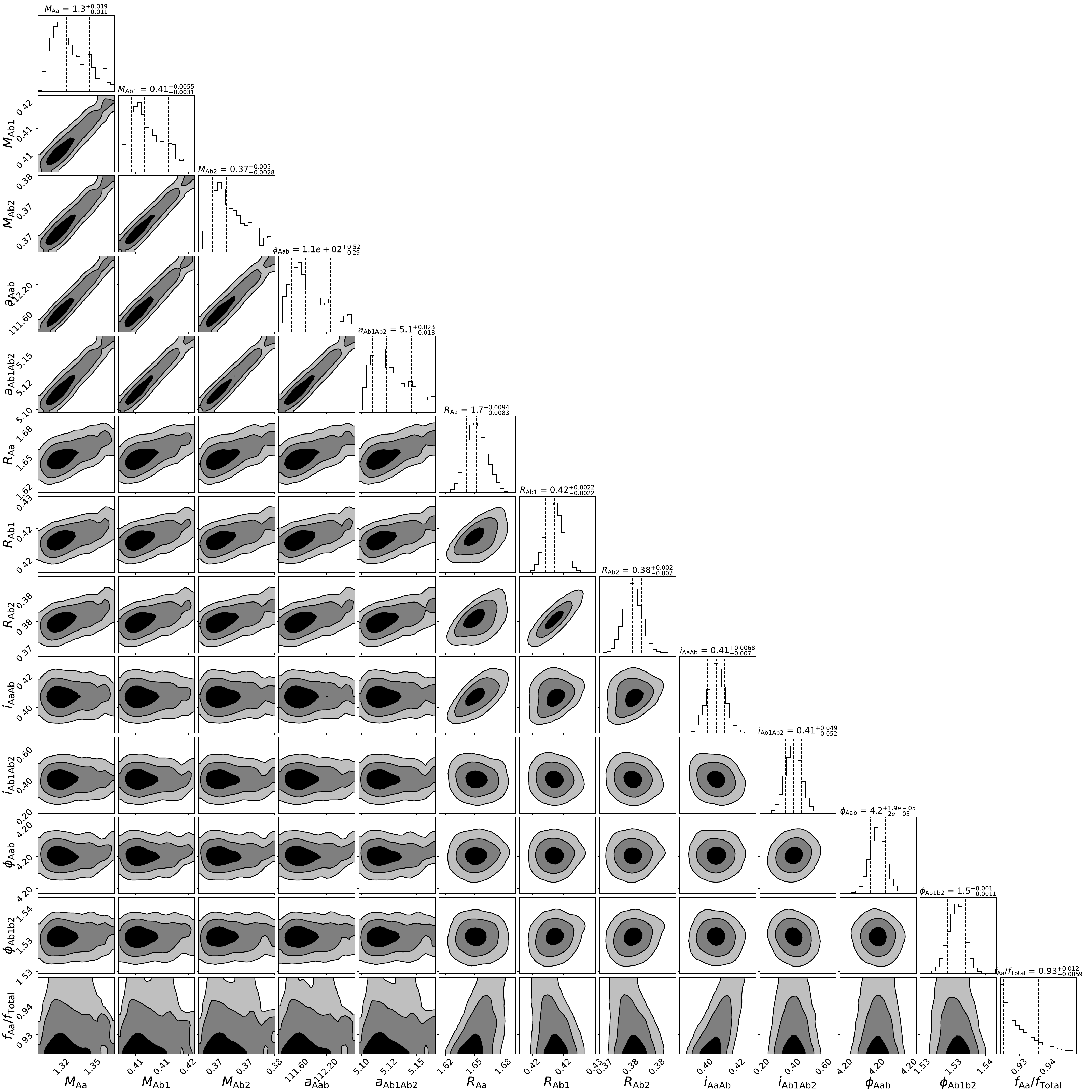}
   \caption{Corner plot showing the distributions of all simulated parameters for the 0.92-0.96 \lfrac\ case. Meaningful constraints can only be placed on the radii, phases, and inclination angles; the masses, separations, and light fraction can be considered unconstrained. Inclination angles are defined with $0^{\circ}$ being edge-on.}
   \label{fig:cornerplot9296}
\end{figure*}

\begin{figure*}
   \centering
   \includegraphics[scale=0.25]{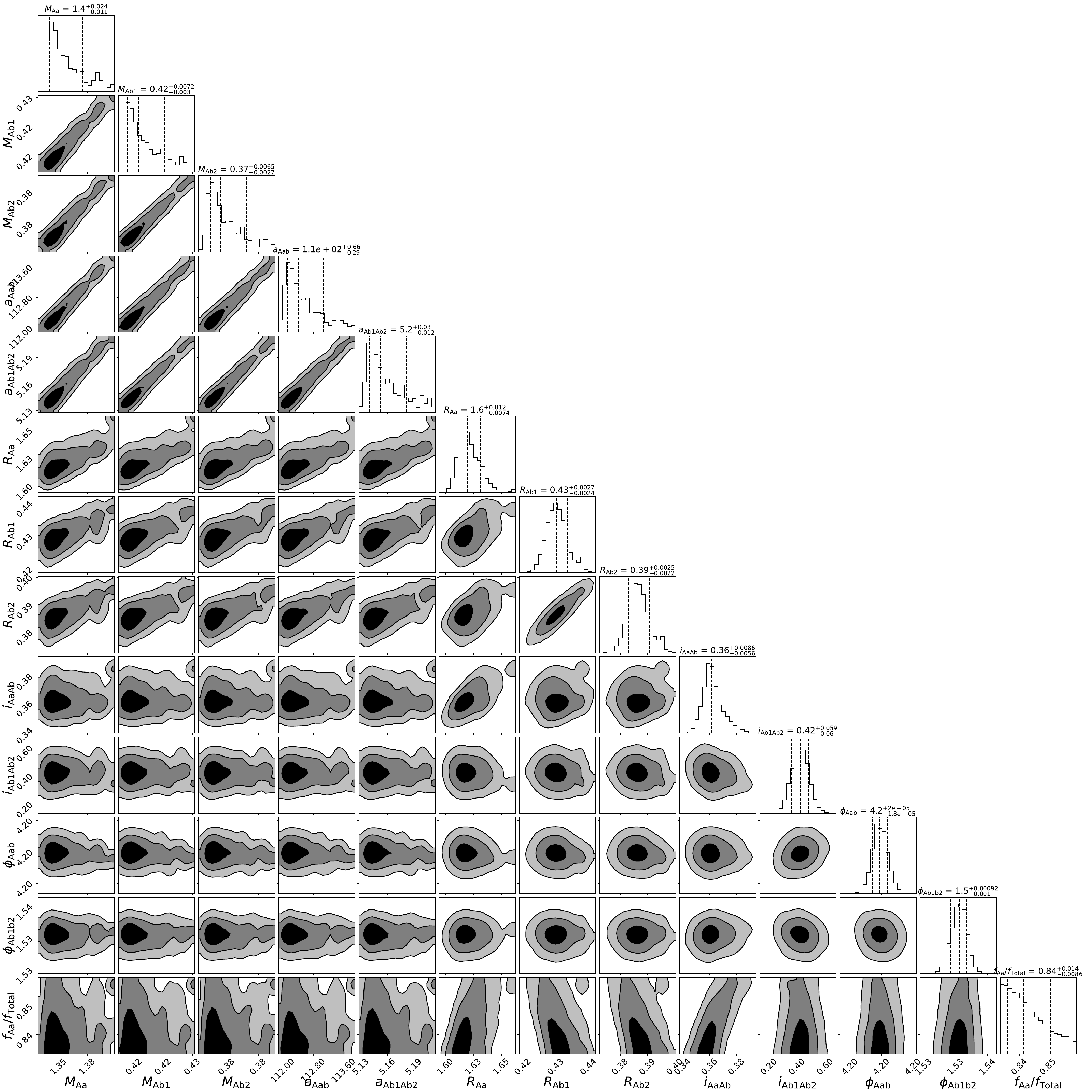}
   \caption{As Fig. \ref{fig:cornerplot9296}, but for 0.83-0.87 \lfrac\ case.}
   \label{fig:cornerplot8387}
\end{figure*}

\begin{figure*}
   \centering
   \includegraphics[scale=0.25]{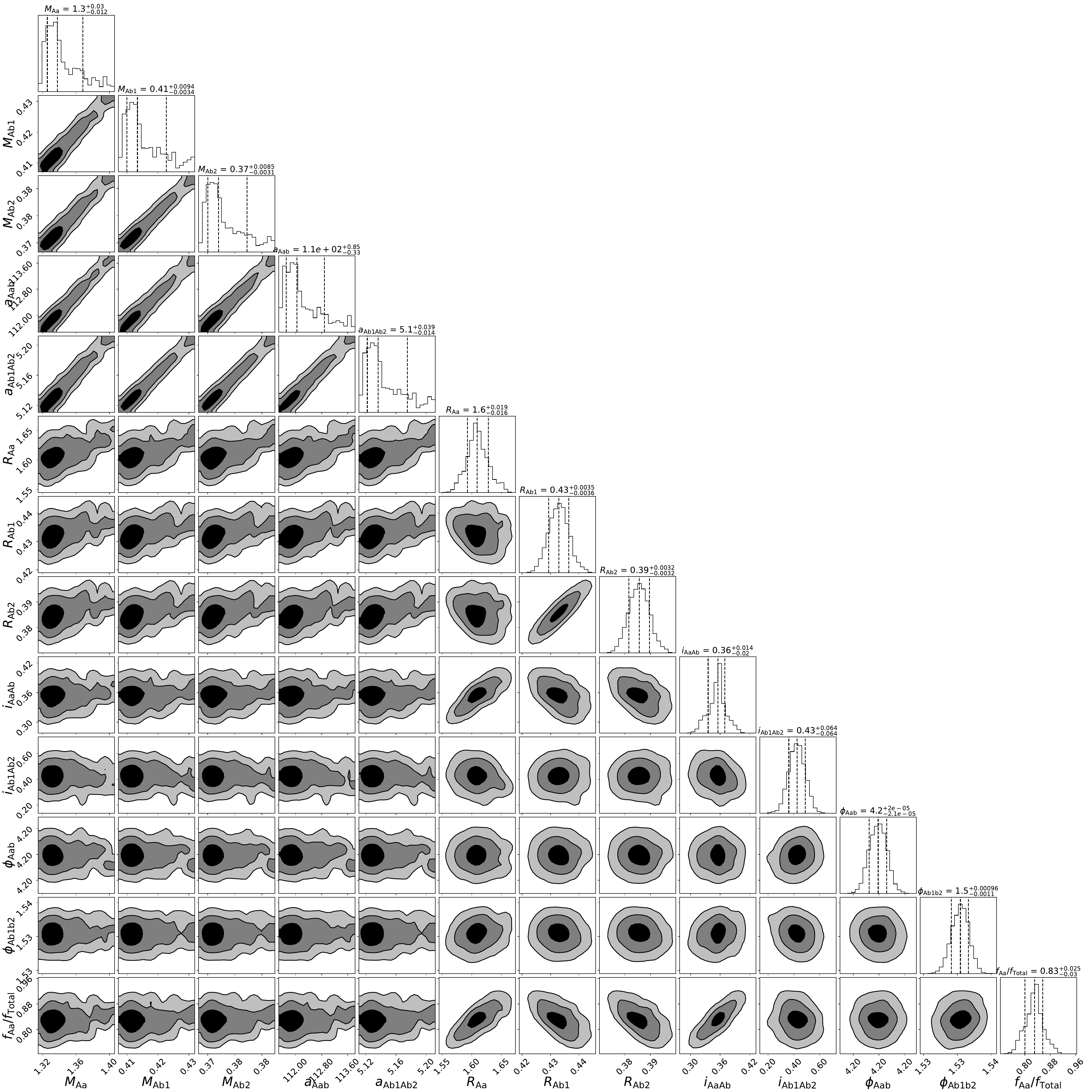}
   \caption{As Fig. \ref{fig:cornerplot9296}, but for the `free' light fraction simulation, where the final state of the 0.83-0.87 \lfrac\ simulation was used to initialise an MCMC simulation with the bounds on light fraction removed.}
   \label{fig:cornerplot8387free}
\end{figure*}

Fig. \ref{fig:cornerplot9296}, \ref{fig:cornerplot8387}, and \ref{fig:cornerplot8387free} present the distributions of each of the physical parameters in our n-body simulations. The mean of each parameter's distribution along with formal 1-$\sigma$ uncertainties are also presented, although it should be noted that not all of these can be considered meaningful. 

\begin{table*}[]
\centering
\begin{tabular}{llllllll}
\lfrac & 0.92-0.96 &  & 0.83-0.87 &  & Free &  & Units \\
\hline
 & Mean & $\sigma$ & Mean & $\sigma$ & Mean & $\sigma$ &  \\
 \hline
$\frac{M_{\rm Aa}}{M_{\rm Ab1}+M_{\rm   Ab2}}$ & 3.60899 & 0.009644 & 3.616071 & 0.011244 & 3.609757 & 0.011643 & - \\
$\frac{M_{\rm Ab1}}{M_{\rm Ab2}}$ & 1.113493 & 0.001189 & 1.113638 & 0.001121 & 1.113805 & 0.001068 & - \\
$\frac{a_{\rm Aab}}{a_{\rm Ab1Ab2}}$ & 21.8186 & 0.011018 & 21.819058 & 0.011902 & 21.814642 & 0.01216 & - \\
$P_{\rm Aab}$ & 94.294862 & 0.000079 & 94.29485 & 0.000086 & 94.294881 & 0.000088 & Days \\
$P_{\rm Ab1Ab2}$ & 1.522248 & 0.000001 & 1.522248 & 0.000001 & 1.522248 & 0.000001 & Days \\
$R_{\rm Aa}$ & 1.651985 & 0.008722 & 1.62128 & 0.009831 & 1.609895 & 0.018195 & R\solar \\
$R_{\rm Ab1}$ & 0.421674 & 0.002214 & 0.432398 & 0.002659 & 0.432924 & 0.003544 & R\solar \\
$R_{\rm Ab2}$ & 0.378438 & 0.002011 & 0.38805 & 0.002384 & 0.388529 & 0.003222 & R\solar \\
$i_{\rm AaAb}$ & 0.408563 & 0.006962 & 0.362485 & 0.007541 & 0.35407 & 0.017894 & $^{\circ}$ \\
$i_{\rm Ab1Ab2}$ & 0.405657 & 0.051257 & 0.42117 & 0.059913 & 0.424795 & 0.063631 & $^{\circ}$ \\
$\phi_{\rm Aab}$ & 4.199443 & 0.00002 & 4.199436 & 0.00002 & 4.199438 & 0.00002 & Radians \\
$\phi_{\rm Ab1b2}$ & 1.532349 & 0.001102 & 1.532399 & 0.000987 & 1.532342 & 0.001016 & Radians
\end{tabular}
\caption{Means and standard deviations for mass ratios, orbital periods, radii, inclination angles, and phases of the triple for each light fraction. We define our inclination angles with $0^{\circ}$ being edge-on. The merit values of the converged state of all three simulations vary between 650-630, 645-625, and 645-625 for the 0.92-0.96, 0.83-0.87, and free light fraction simulations respectively. This implies a small statistical benefit when considering light fractions below 0.92, but no statistical benefit to the fit when considering light fractions below 0.83.}
\label{tab:eclipseall}
\end{table*}

We conclude based on our MCMC simulations that our eclipse modelling is unable to provide useful constraints on the light fraction. Based purely on the eclipse modelling, there is a very small statistical improvement for a lower light-fraction similar to that put forth in \cite{heliminiak2017} ($\approx0.85)$ over the higher light fraction that is implied from our spectroscopic analysis. This is reflected in the \lfrac\ distributions for each bounded light fraction simulation, with both favouring their lower bound. Our third simulation, where the light fraction is unbounded, quantifies this preference, with the walkers covering a significantly wider span of the parameter space (roughly between 0.75-0.9, with a formal distribution characterised by mean of 0.83 and a 0.03 $1-\sigma$ uncertainty). Despite this, there is no statistical improvement to the merit function compared to the 0.83-0.87 bounding case.

The statistical improvement of the lower light fraction is so small that we speculate that it could be eliminated by small changes to the eclipse model (such as the inclusion of non-spherical geometry or a using different limb-darkening law) or data reduction processes could eliminate or otherwise affect this improvement. Comparatively, the spectral modelling of the disentangled spectra responds far more strongly (and directly) to the light fraction. It is for this reason that despite the apparently well defined distribution for the light fraction seen in Fig. \ref{fig:cornerplot8387free}, we assert that the small statistical improvement we see in our eclipse modelling is far less significant than the relatively strong dependence we see in our spectral modelling (see Section \ref{sec:res:spec}). Nonetheless, we consider the effect of uncertainty in the light fraction on the other physical parameters of the system.

\begin{figure*}
   \centering
   \includegraphics[scale=0.35]{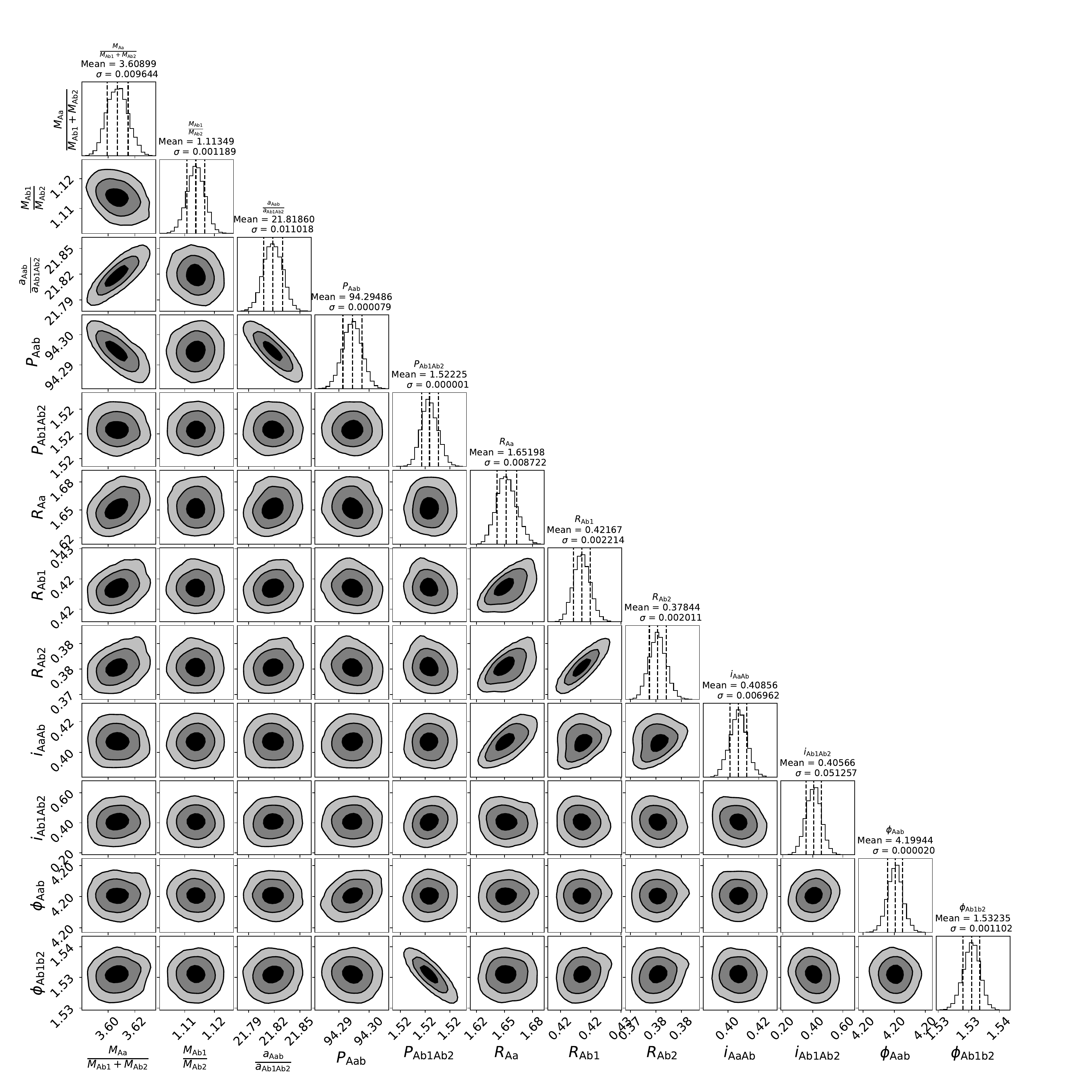}
   \caption{Corner plot showing the distributions system parameters with meaningful constraints for the 0.92-0.96 \lfrac\ case.}
   \label{fig:cornerproppaplot9296}
\end{figure*}

\begin{figure*}
   \centering
   \includegraphics[scale=0.35]{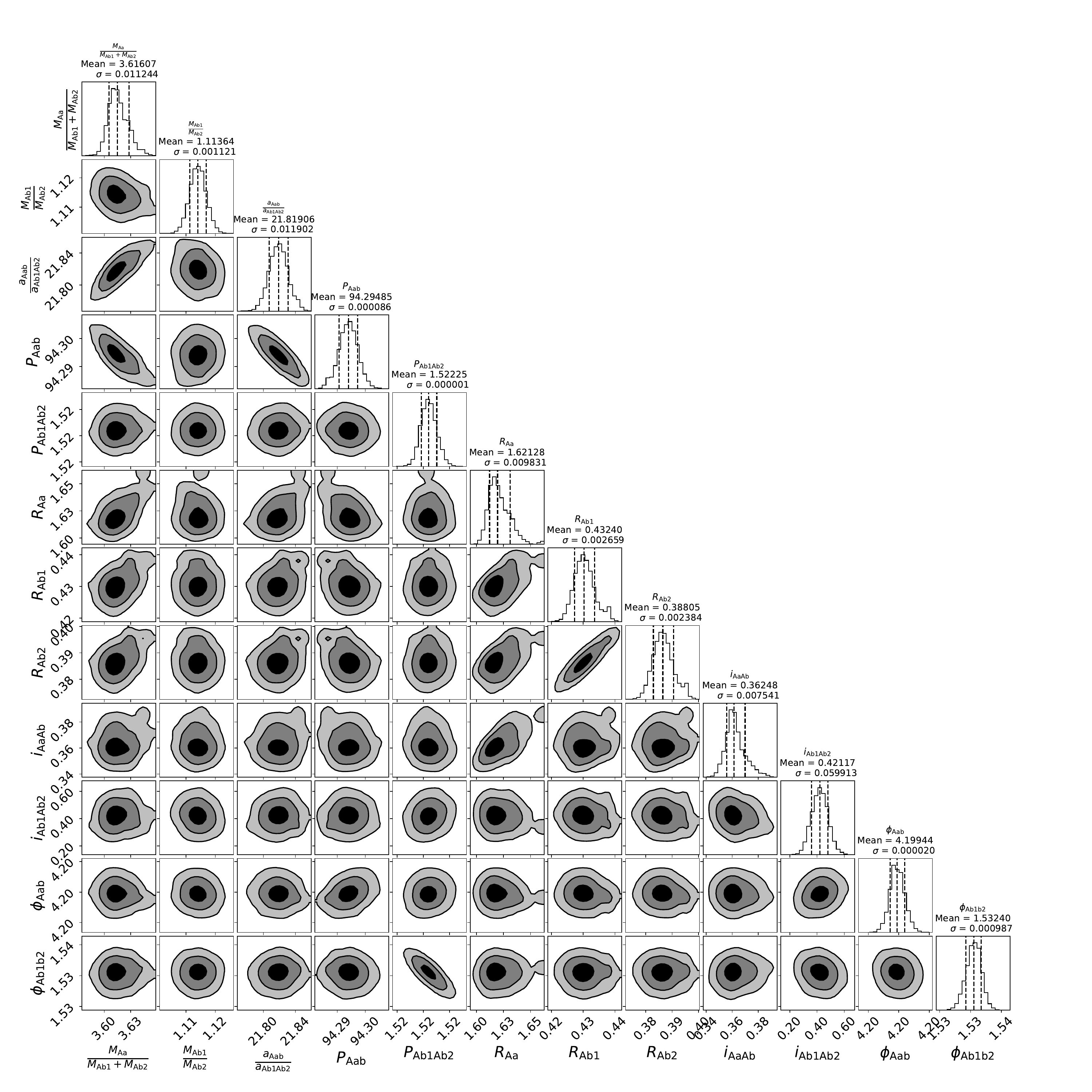}
   \caption{Corner plot showing the distributions system parameters with meaningful constraints for the 0.83-0.87 \lfrac\ case.}
   \label{fig:cornerproppaplot8387}
\end{figure*}

\begin{figure*}
   \centering
   \includegraphics[scale=0.35]{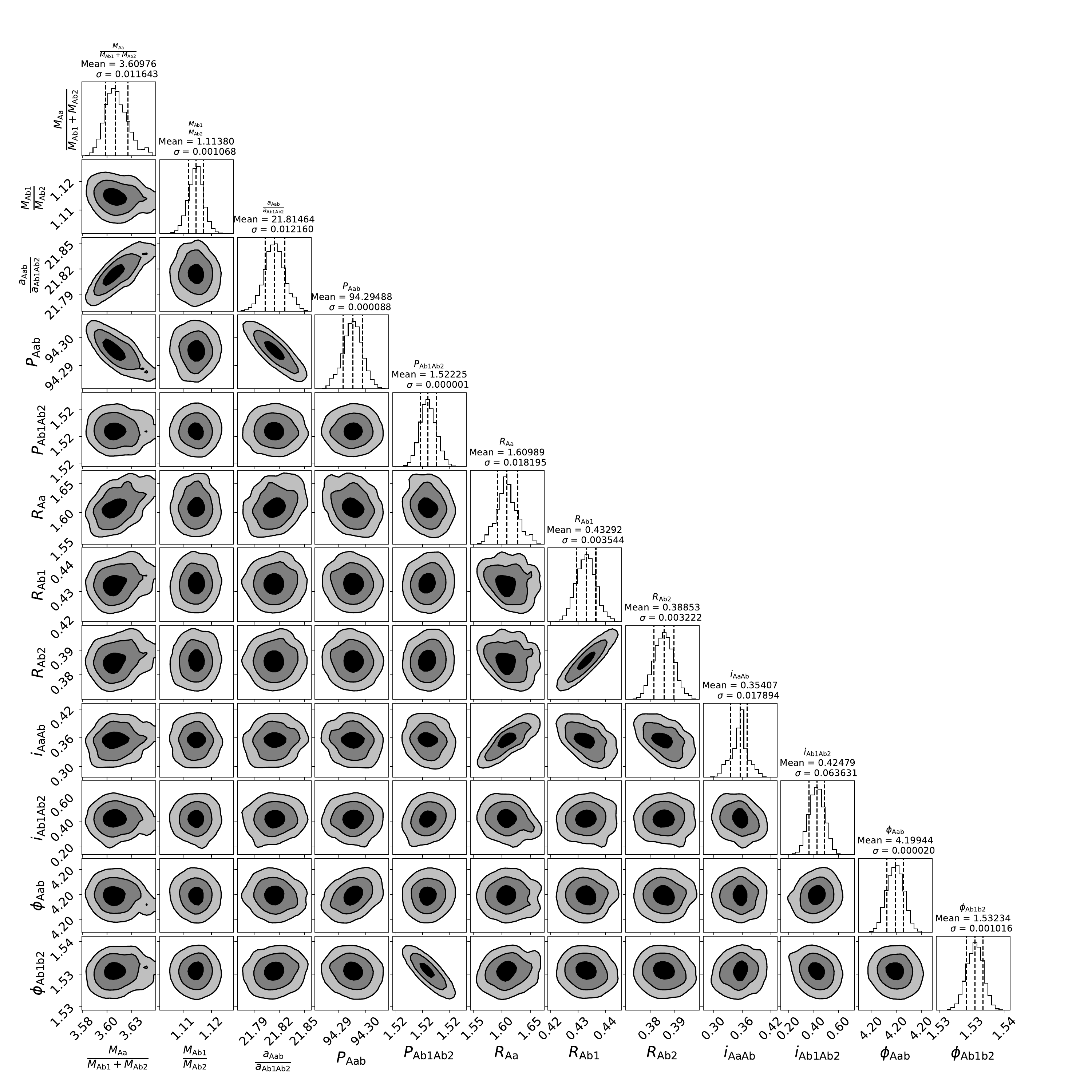}
   \caption{Corner plot showing the distributions system parameters with meaningful constraints for the `free' \lfrac\ case.}
   \label{fig:cornerproppaplot8387free}
\end{figure*}

The meaningful quantitative results for each simulation are reported in Table \ref{tab:eclipseall} and their distributions are shown in Fig. \ref{fig:cornerproppaplot9296}, \ref{fig:cornerproppaplot8387}, and  \ref{fig:cornerproppaplot8387free}. This includes both directly modelled parameters such as the stellar radii and indirectly modelled parameters such as orbital periods and mass ratios.

The orbital period for the A binary is worthy of particular attention. Due to the complicated eclipse geometry, obtaining precise measurements of the orbital period is difficult using traditional eclipse-timing methods. Working in Fourier space does not circumvent this issue; the complex geometry manifests as significant variations in the spacings between orbital harmonics. When examining the distribution of the spacings between the 94.2d orbital harmonics, a broad lop-sided bimodal distribution results. An orbital period derived from the RV curve avoids this issue, but will not be as precise due to both the naturally higher scatter in RV measurements and the far lower number of RV measurements when compared to photometry. One benefit of our physical model is that we can compute the orbital period of the center of mass of the Ab binary with Aa, thereby benefiting from the large number of precise photometry measurements while not relying on reliable eclipse timings.

This results in an orbital period of of 94.29486d, precise to slightly better than \tento{-4}d. This compares with orbital periods of 94.1982(7)d from the \kepler\ eclipsing binary catalogue (KEBC) \citep{prsa2011,slawson2011,orosz2015}, $94.09\pm0.11$d from \cite{shibahashi2012}, $94.44\pm0.24$d from \cite{balona2014}, and our own radial velocity analysis of $94.10\pm0.11$d. \cite{heliminiak2017} hold the orbital period fixed at 94.226d in their analysis. The variation in the literature values reflects the difficulty in determining an accurate orbital period for this system. Comparing with these previous findings, our orbital period is on the higher end (excepting \cite{balona2014}'s value, which has a very high reported uncertainty). Considering the varying eclipse geometry of the system, a value of around 94.2d is likely to be sufficiently precise for most applications.

We arrive at an even more precise constraint on the orbital period of the Ab binary, $1.522248\pm$\tento{-6}d. 
This compares to 1.5222786(19)d from the \kepler\ eclipsing binary catalogue (KEBC) \citep{prsa2011,slawson2011,orosz2015}, and 1.5222468(25)d from \cite{heliminiak2017}. This is an interesting comparison, as we are \textit{not} computing this based on the eclipses of the Ab binary directly, but instead only on the Ab binary's transits of Aa. There are far fewer examples of the 94.2d eclipses, but the signal of each is much higher than the faint 1.52d eclipses of the Ab binary (see Fig. \ref{fig:pofourphasefold}). In order to identify orbital harmonics as part of our frequency analysis, we make use of \cite{heliminiak2017}'s orbital period, which works perfectly well, giving us high confidence in their orbital period as a sufficiently accurate solution. Our period of $1.522248$d also provides a satisfactory phase-fold and sufficient predictive power to find the orbital harmonics in the Lomb-Scargle periodogram. Noting that our solution lies within the formal uncertainty of \cite{heliminiak2017}, we conclude that our solution is accurate to a useful degree.

We also perform a sanity check of the high formal precision by considering the maximum allowable phase offset from the `true' orbital period. Fig. \ref{fig:esc9296} reveals an excellent fit to the data; a discrepancy in orbital period of only so much, then, can be permitted before the Ab binary becomes too out-of-phase by the end of a simulation. There is, at worst, a discrepancy of three minutes (a tenth of a \kepler\ timestamp) between our predictions and data at these extreme ends of the simulation, and we'll assume that our model is perfectly in phase with the `true' system halfway between our first and last eclipse. Our last eclipse ends at 1515 days, and our first begins at 195 days. A 1.5222d day period binary will orbit 867 times during this time period, or around 433 times from the time of hypothetical perfect alignment with reality. Then our computed orbital period would be in error with truth by $3\times60/433=0.4$ seconds, or \timestento{4.63}{-6}d. This can be regarded as an upper bound; the wings of our eclipses appear to be in agreement by better than a tenth of a \kepler\ time stamp at each extreme end of the \kepler\ cycle, and the case of perfect orbital phase alignment with reality is the worst-case scenario when considering how much of a constraint we can place on our precision. This leads us to the conclusion that our formal precision of \tento{-6}d on the orbital period for the Ab binary is plausible. We speculate that this high precision may be useful for the purposes of scheduling observations from the ground, but note that the \cite{heliminiak2017} solution for this orbital period is sufficiently accurate and precise for the purposes of harmonic identification.

The separation ratio is determined to be $21.81\pm 0.01$, and insensitive to the light fraction. We could not find other estimates for this quantity reported in previous works, but it has relevance for the stability of the system, which we discuss later in this section.

We find the ratio of the mass of Aa to the total mass of the Ab binary, $\frac{M_{\rm Aa}}{M_{\rm Ab1}+M_{\rm Ab2}}$, to be $3.61\pm0.01$. This is significantly higher than the mass ratio of less than 2 implied by \cite{heliminiak2017}'s combination of isochrone fitting and radial velocity measurements of the F-type star. The mass ratio of the Ab binary, $\frac{M_{\rm Ab1}}{M_{\rm Ab2}}$, we find to be $1.113\pm0.001$, a closer mass ratio than that implied by \cite{heliminiak2017}'s isochrone estimate of 1.372 but not significantly inconsistent in the context of the \cite{heliminiak2017} estimate's high uncertainty, which we conservatively estimate as between 0.6 and 1.7 by taking the upper and lower uncertainty limits on the masses of each component, and vice versa.

Returning to directly modelled parameters, the radius of Aa is estimated at $1.64\pm0.06$~R\solar\ when considering the variation between all three MCMC simulations. This parameter is most directly linked to the assumption surrounding light fraction; $1.62\pm0.01$~R\solar\ for 0.83-0.87, $1.65\pm0.01$~R\solar\ for 0.92-0.96, and $1.60\pm0.02$~R\solar\ for the free light fraction case. \cite{heliminiak2017} estimate the radius of Aa to be smaller, 1.376(13) R\solar, based on their isochrone fitting.

\cite{heliminiak2017}'s isochrone fitting process was applied to the B binary; the age of this and the K-magnitude was then used to infer the properties of Aa. These inferred properties include a temperature of 8440(240)K and a log$(g)$ of 4.38(1). This is a considerably hotter and more compact star than what spectroscopy from both \cite{niemczura2015} and this work implies, a fact also noted by \cite{heliminiak2017}. All isochrone fits have a large degree of modelling uncertainty inherited from the underlying stellar models. In contrast, our eclipse modelling provides a direct, almost model-independent probe of stellar radii, albeit with some degeneracy with \lfrac. We therefore consider the larger radius for Aa from our eclipse modelling to be a better solution than the radius derived from \cite{heliminiak2017}'s isochrone fits.

Our estimated radii for Ab1 and Ab2 are  $0.42\pm0.01$~R\solar\ and $0.38\pm0.01$~R\solar, with most of this reported uncertainty again stemming from our estimate of the impact of light fraction assumptions. These radii are smaller than those implied by \cite{heliminiak2017}'s isochrone fits of $0.61\pm0.1$~R\solar\ and $0.52\pm0.1$~R\solar, which they derive using their obtained masses and the implied 35 Myr age (placing these stars on the pre-main sequence) of the system from their isochrone fits to the B binary. By the same logic, our different radii could imply a different system age, but this discrepancy could also simply be due to the high degree of modelling uncertainty associated with isochrone fitting.

The inclination angle\footnote{We define our inclination angles with $0^{\circ}$ being edge-on.} \iAaAb\ is somewhat sensitive to the light fraction, varying between $0.36^{\circ}$ and $0.41 ^{\circ}$, with a formal uncertainty less than $0.01^{\circ}$ for the 0.83-0.87 and 0.92-0.96 light fraction cases. The inclination angle \iAbonebtwo\ is insensitive to light fraction, but also less precise; both simulations converge to the same result within uncertainty, $0.41\pm0.06^{\circ}$. This lower precision is expected to be sensitive to the wings of the eclipses, but mostly to the details of the flux spike in the super-eclipses caused by the simultaneous eclipse of Ab1, Ab2, and Aa. Although this spike occurs so briefly that it is poorly resolved in \kepler\ long cadence data, comparison with short-cadence \kepler\ data (see Fig. \ref{fig:esc9296sc}) reveals that the spike is actually well-reproduced in all cases. This implies that we may be overestimating the uncertainty for this inclination angle, but that the mean value is likely accurate.

The phase terms (see Fig. \ref{fig:triple_diagram}), for similar reasons to the orbital period of the Ab binary, are also constrained to high precision, and are insensitive to the assumed light fraction. We compute phase terms of \fAab=$4.19944\pm 0.00002$~rad and \fAbonebtwo=$1.532\pm0.001$~rad for the Aa-Ab and Ab1-Ab2 binaries, respectively. The higher precision on \fAab\ may seem counter-intuitive given that we could constrain the orbital period more precisely for the Ab binary. However, this can be understood as a natural consequence of the larger lever arm in the Aa-Ab binary; a small change in \fAab\ results in a larger change in the position of the center of mass of the Ab binary, and therefore the timing of its eclipses.

Finally, a test of the long-term stability of the A triple was performed by calculating the critical ratio of the minimum periastron distance of the outer orbit to the maximum apastron distance of the inner orbit according to Equation 2 from \cite{eggleton1995}. This can be done using the outer mass ratio $\frac{M_{\rm Aa}}{M_{\rm Ab1}+M_{\rm Ab2}}$, inner mass ratio $\frac{M_{\rm Ab1}}{M_{\rm Ab2}}$, separation ratio $\frac{a_{\rm Aab}}{a_{Ab1Ab2}}$, and eccentricities of the inner and outer binary. More complex stability criteria exist accounting for differences in the mutual inclination between the outer and inner binary \citep[see, for example,][] {vynatheya2022}. However, as both of the Aa-Ab binary and the Ab binary are nearly edge-on, these effects can be safely neglected. We find that the A triple is very stable, with a ratio of 21.82 and a critical ratio of 7.95.

\subsection{Spectroscopic analysis}
\label{sec:res:spec}

\begin{figure}
   \centering
   \includegraphics[scale=0.6]{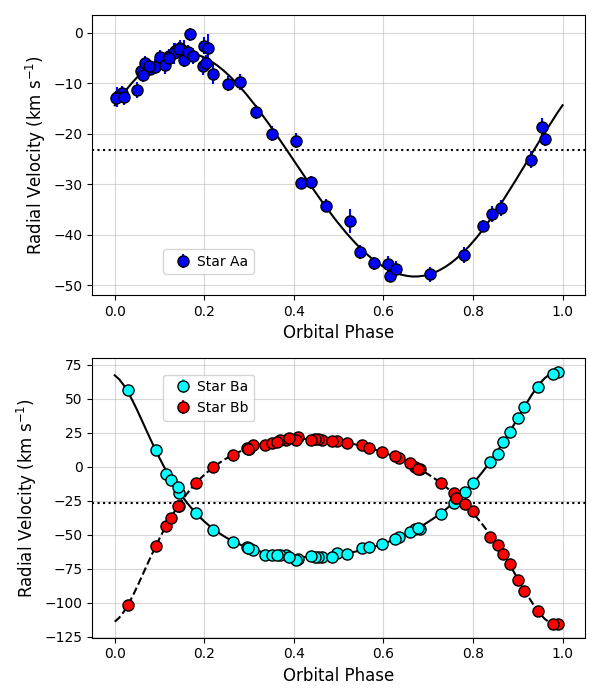}
   \caption{Spectroscopic orbits for the main star in KIC 4150611 (Aa), and for the pair of G stars (Ba+Bb). The dotted lines represent the center-of-mass velocities of each system.}
   \label{fig:rvs_torres}
\end{figure}

Orbital solutions based purely on our radial velocity measurements for the single-lined binary (Aa) and the double-lined binary (Ba+Bb) are shown in Fig. \ref{fig:rvs_torres}, with orbital parameters reported in Table \ref{tab:spectro_torres}. Comparing with the analogous Table 3 from \cite{heliminiak2017}, we find lower values of $a \sin i$ but otherwise the properties of the system are consistent. A non-zero eccentricity is found for the Aa system, but it is very small ($0.032 \pm 0.015$); the binary is circular at the 2-$\sigma$ level.

We also derived the spectroscopic flux ratios between the components at the mean wavelength of our observations using \texttt{TRICOR}. We obtained $f_{\rm Ab1}/f_{\rm Aa} = 0.0365 \pm 0.0035$ and $f_{\rm Ab2}/f_{\rm Aa} = 0.0355 \pm 0.038$. After transforming these to the \kepler\ band using synthetic spectra based on PHOENIX model atmospheres \citep{husser2013}, we derived a fractional light contribution for Aa of $f_{\rm Aa}/f_{\rm Total} = 0.91 \pm 0.02$.

\begin{table}[]
\begin{tabular}{llll}
 & Aa (SB1) & Ba + Bb (SB2) & Units \\ \cline{2-4} 
\multicolumn{1}{l|}{$P$} & $94.10 \pm 0.11$ & $8.653243 \pm 0.000073$ & Days \\
\multicolumn{1}{l|}{$T_{\rm peri}$ } & $56027.4 \pm 7.8$ & $56050.3738 \pm 0.0067$ & Days \\
\multicolumn{1}{l|}{$K_1$} & $22.19 \pm 0.34$ & $67.72 \pm 0.25$ & \kms \\
\multicolumn{1}{l|}{$K_2$} & - & $68.09 \pm 0.16$ & \kms \\
\multicolumn{1}{l|}{$\gamma$} & $-26.45 \pm 0.25$ & $23.203 \pm 0.077$ & \kms \\
\multicolumn{1}{l|}{$e$} & $0.032 \pm 0.015$ & $0.3701 \pm 0.0015$ & - \\
\multicolumn{1}{l|}{$\omega_1$} & $302 \pm 30$ & $13.18 \pm 0.31$ & $^\circ$ \\
\multicolumn{1}{l|}{$f(m)$} & $0.1063 \pm 0.0049$ & - & M\solar \\
\multicolumn{1}{l|}{$M_1 \sin^3 i$} & - & $0.9029 \pm 0.0052$ & M\solar \\
\multicolumn{1}{l|}{$M_2 \sin^3 i$} & - & $0.8980 \pm 0.0066$ & M\solar \\
\multicolumn{1}{l|}{$a_1 \sin i$} & $28.69 \pm 0.44$ & $7.486 \pm 0.026$ & $10^6$ km \\
\multicolumn{1}{l|}{$a_2 \sin i$} & - & $7.527 \pm 0.017$ & $10^6$ km \\
\multicolumn{1}{l|}{$\sigma_1$} & 1.56 & 1.08 & \kms \\
\multicolumn{1}{l|}{$\sigma_2$} & - & 0.66 & \kms
\end{tabular}

    \caption{Orbital properties of Aa, Ba, and Bb directly from the radial velocity measurements, analogous to Table 4 from \cite{heliminiak2017}. $T_{\rm peri}$ is measured relative to $\rm BJD = 2$,400,000; that is, $T_{\rm peri} = \rm BJD - 2$,400,000. Compared to the previous analysis by \cite{heliminiak2017}, we obtain lower values of $a \sin i$ for all three components; other properties are similar.}
\label{tab:spectro_torres}
\end{table}

\begin{table}[]
\begin{tabular}{llll}
 & Aa (SB1) & Ba + Bb (SB2) & Units \\ \cline{2-4} 
\multicolumn{1}{l|}{$P$} & $94.07 \pm 0.11$ & $8.653290 \pm 0.000075$ & Days \\
\multicolumn{1}{l|}{$T_{\rm peri}$} & $56024.1 \pm 6.7$ & $56050.3728 \pm 0.0065$ & Days \\
\multicolumn{1}{l|}{$K_1$} & $22.11 \pm 0.36$ & $67.86 \pm 0.26$ & \kms \\
\multicolumn{1}{l|}{$K_2$} & - & $68.26 \pm 0.18$ & \kms \\
\multicolumn{1}{l|}{$\gamma$} & $-26.34 \pm 0.36$ & $-23.203 \pm 0.082$ & \kms \\
\multicolumn{1}{l|}{$e$} & $0.037 \pm 0.016$ & $0.3713 \pm 0.0017$ & - \\
\multicolumn{1}{l|}{$\omega_1$} & $289 \pm 26$ & $13.14 \pm 0.31$ & $^{\circ}$ \\
\multicolumn{1}{l|}{$f(m)$} & $0.1052 \pm 0.0051$ & - & M\solar \\
\multicolumn{1}{l|}{$M_1 \sin^3 i$} & - & $0.9076 \pm 0.0056$ & M\solar \\
\multicolumn{1}{l|}{$M_2 \sin^3 i$} & - & $0.9023 \pm 0.0070$ & M\solar \\
\multicolumn{1}{l|}{$a_1 \sin i$} & $28.59 \pm 0.51$ & $7.497 \pm 0.027$ & $10^6$ km \\
\multicolumn{1}{l|}{$a_2 \sin i$} & - & $7.541 \pm 0.018$ & $10^6$ km \\
\multicolumn{1}{l|}{$\sigma_1$} & 1.545 & 1.05 & \kms \\
\multicolumn{1}{l|}{$\sigma_2$} & - & 0.68 & \kms
\end{tabular}
\caption{As Table \ref{tab:spectro_torres}, but from the disentangled residual spectra. The values are consistent with Table \ref{tab:spectro_torres} within uncertainty.}
\label{tab:spectro_kresimir}
\end{table}


During spectral disentangling, the orbital parameters of the binary or hierarchical-multiple system are optimised simultaneously with the calculations of the components' spectra. The orbital parameters optimised during our Fourier disentangling check (which did not fix radial velocities) are shown in Table \ref{tab:spectro_kresimir}, and are identical to the results in Table \ref{tab:spectro_torres} within the uncertainties. A determination of the light fraction is also naturally obtained through the spectral disentangling process. We find that the two G stars contribute 3-4\% each to the spectrum, consistent with the spectroscopic flux-ratio analysis.

\begin{table}[]
\centering
\begin{tabular}{llll}
Variable & Value & $\sigma$ & Units \\
\hline
$T_{\rm   eff}$ & 7280 & 70 & K \\
log$(g)$ & 4.14 & 0.18 & dex \\
$v_{\rm micro}$ & 3.61 & 0.19 & \kms \\
$v \sin i$ & 127 & 4 & \kms \\
{[}M/H{]} & -0.23 & 0.06 & dex \\
\lfrac & 0.94 & 0.01 & -
\end{tabular}
\caption{Stellar parameters from the optimal \gssp\ fit to Aa. The fit to the disentangled spectrum of Aa is shown in Fig. \ref{fig:GSSPfitAa}}
\label{tab:GSSPfitAa1b}
\end{table}

\begin{table}[]
\centering
\begin{tabular}{llll}
Variable & Value & $\sigma$ & Units \\
\hline
$T_{\rm   eff}$ & 5620 & 92 & K \\
log$(g)$ & 4.41 & 0.17 & dex \\
$v_{\rm micro}$ & 2 & (fixed) & km/s \\
$v_{\rm sin(i)}$ & 8.9 & 1.1 & km/s \\
{[}M/H{]} & -0.18 & 0.12 & dex \\
\lfracBa & 0.038 & 0.003 & -
\end{tabular}
\caption{Stellar parameters from the optimal \gssp\ fit to Ba, the brighter of the two G stars in the B binary. The fit to the disentangled spectrum of Aa is shown in Fig. \ref{fig:GSSPfitBa}}
\label{tab:GSSPfitBa}
\end{table}

\begin{table}[]
\centering
\begin{tabular}{llll}
Variable & Value & $\sigma$ & Units \\
\hline
$T_{\rm   eff}$ & 5580 & 65 & K \\
log$(g)$ & 4.42 & 0.15 & dex \\
$v_{\rm micro}$ & 2 & (fixed) & km/s \\
$v_{\rm sin(i)}$ & 9.2 & 1.1 & km/s \\
{[}M/H{]} & -0.19 & 0.1 & dex \\
\lfracBb & 0.035 & 0.003 & -
\end{tabular}
\caption{Stellar parameters from the optimal \gssp\ fit to Bb, the dimmer of the two G stars in the B binary. The fit to the disentangled spectrum of Ab is shown in Fig. \ref{fig:GSSPfitBb}}
\label{tab:GSSPfitBb}
\end{table}

The disentangled spectra of the Aa, Ba, and Bb components of the system are analysed with the Grid Search in Stellar Parameters \citep[\gssp,][]{tkachenko2015} software package. This package employs atmospheric models and spectral synthesis to analyse stellar spectra of single and binary stars taken in an arbitrary wavelength range and with arbitrary resolving power. \gssp\ uses a pre-computed grid of atmosphere models \citep[computed using with the \texttt{LLmodels} software package by][]{shulyak2004} and the \texttt{SynthV} radiative transfer code \citep{tsymbal1996} for spectral synthesis. Atmosphere models and synthetic spectra are computed under the assumption of local thermodynamic equilibrium (LTE) and plane-parallel geometry, and all quoted uncertainties are within the scope of these assumptions, and as such can be considered lower bounds on the true uncertainty.

The optimal \gssp\ model fit to the disentangled spectrum for the Aa primary component is shown in Fig. \ref{fig:GSSPfitAa}; the corresponding best fitting parameters are listed in Table \ref{tab:GSSPfitAa1b}. We confirm previous findings \citep{niemczura2015} that the primary component is a main-sequence F-type star with a moderate rotation rate and a somewhat sub-solar metallicity. The effective temperature ($7280\pm70$~K), surface gravity ($4.14 \pm 0.18$), and metallicity ($-0.23 \pm 0.06$~dex) we obtain from our best fitting spectral model are fairly consistent with the values obtained by \cite{niemczura2015} ($T_{\rm eff}\approx 7400$~K, log$(g)\approx3.8$, and {[}M/H{]} $\approx -0.16$~dex). It should be noted that \cite{niemczura2015} analyzed KIC~4150611 as if it were single, without consideration for the roughly 6\% contribution from the B binary. For this reason, our atmospheric analysis can be considered an improvement on the existing literature values. Aa's surface properties are consistent with the statistical characteristics of the sample of $\gamma$-Dor pulsators presented in \cite{mombarg2024} and \cite{fritzewski2024}, lending support to previous assertions of the star being the source of the pulsations present in the \kepler\ light curve.

The GSSP solution suggests that the light contribution of the Aa component constitutes $94\pm1$\% to the total light of the system. This result is at odds with the light fraction determined by \cite{heliminiak2017} (0.85 to high precision) and - tenuous - light fraction determination from our eclipse modelling when considering the free light-fraction case ($0.83\pm0.03$). However, modelling of the disentangled spectrum has a more direct and stronger dependency on the light fraction. As a consistency check, we also perform the analysis of the disentangled spectrum with the light fraction fixed to 0.85. This lower light fraction pushes the star into the regime of late A-type stars and the quality of the obtained fit to the disentangled spectrum is significantly inferior to the solution reported in Table \ref{tab:GSSPfitAa1b}. \kepler\ and \textit{TRES} have similar response functions (spanning 420-900~nm and 380-890~nm, respectively), making wavelength-dependent instrumental effects unlikely as the source of the discrepancy in light fraction. However, we note that the disentangling process is conducted using only the shorter-wavelength section of the pass-band. We estimate that the variation in light fraction across the \kepler\ response function is at most 4.2\% (at the longest wavelengths), leading to an estimate of at most 1-2\% discrepancy between the short-wavelength regime and the entire response function. As even the maximum discrepancy would be insufficient to explain the difference in the preferred light fractions in the two methods, it seems extremely unlikely that the two can be reconciled by instrumental effects.

Ultimately, we are able to build a satisfactory eclipse model and spectral fit when considering a light fraction between 0.92 and 0.94, values consistent with both the disentangling and \texttt{TRICOR} flux ratio analysis. When considering a lower light fraction, we are unable to construct a satisfactory fit to the disentangled spectrum. An assessment of the possible impact of instrumental effects between \kepler\ and \textit{TRES} implies that such effects could account for at most 1-2\% discrepancy in the observed light fraction. As such, we consider it fair to compare the light fraction derived through spectrosopy with \kepler\ data. This in turn leads us to conclude that the 0.92-0.96 light fraction eclipse models are the preferred set of solutions. None the less, a conservative estimate of the parameter uncertainty should consider that the results from all three model sets.

The atmospheric properties of the two G stars, Ba and Bb, are presented in Tables \ref{tab:GSSPfitBa} and \ref{tab:GSSPfitBb} respectively. For these fits, $v_{\rm micro}$ was fixed at 2 km/s, but other parameters were left free. As noted by \cite{heliminiak2017}, the two G stars appear almost identical, with all of their atmospheric properties having overlapping 1-$\sigma$ errors. The fits to these spectra are shown in Figure \ref{fig:GSSPfitBa} and \ref{fig:GSSPfitBb}, respectively.

Finally, an attempt was made to obtain a radial velocity curve of the two dwarfs Ab1 and Ab2 from the time series of spectra. Such a curve would lead to a mass determination for Aa, Ab1, and Ab2. This attempt was unsuccessful; Ab1 and Ab2 are too dim to be detected in the presence of Aa. If it is possible at all, a dedicated spectroscopic campaign with extremely high signal-to-noise measurements would be required. These observations would preferably be in the near-infrared, where the contrast with the K/M stars will be more favourable. 

\section{Conclusions}

In this work, we present the results of photometric eclipse modelling of the triple component of KIC~4150611 and spectroscopic analysis of the SB3 spectra of the F-type primary Aa and the G+G binary (Ba + Bb). Obtaining constraints on the stellar properties of the F1V primary Aa is of particular value, as this star is a hybrid $\delta$-Scuti/$\gamma$-Dor pulsator exhibiting both high amplitude p-modes and a g-mode period-spacing pattern; an asteroseismic analysis of its pulsations is in preparation.

Our eclipse modelling employed the novel combination of a full n-body model for the triple, with its dynamical evolution handled by the \texttt{Rebound} n-body integrator and the photometry handled by the exo-moon transit model from \texttt{Pandora}. This methodology proved successful in modelling the complex eclipse geometry of the A triple, which varies significantly from eclipse to eclipse depending on the orbital phase of the eclipsing Ab binary. 

Our eclipse modelling places robust constraints on the mass ratios ($\frac{M_{\rm Aa}}{M_{\rm Ab1}+M_{\rm   Ab2}}=3.61\pm0.01$, $\frac{M_{\rm Ab1}}{M_{\rm Ab2}} = 1.113\pm0.001$), the separation ratio ($\frac{a_{\rm Aab}}{a_{\rm Ab1Ab2}} = 21.81\pm 0.01$), orbital periods ($P_{\rm Aab} = 94.29486\pm0.00008$d, $P_{\rm Ab1Ab2} = 1.522248\pm0.000001$d), stellar radii (\RAa\ $ = 1.64\pm0.06$~R\solar, \RAbone\ $ = 0.42\pm0.01$~R\solar, \RAbtwo\ $ = 0.38\pm0.01$~R\solar), inclination angles (\iAaAb\ $ = 0.38\pm0.02^{\circ}$, \iAbonebtwo\ $ = 0.41\pm0.06^{\circ}$, where $0^{\circ}$ is edge on), and orbital phase terms (\fAab=$4.19944\pm 0.00002$~rad, \fAbonebtwo=$1.532\pm0.001$~rad) of the triple, which we calculated to be stable on long time scales. Absolute masses and separations for Aa and its dwarf companions Ab1 and Ab2 could not be obtained; their acquisition requires radial velocity measurements of the very faint Ab binary, which would require a dedicated spectroscopic campaign of its own.

Our spectroscopic analysis made use of 51 \textit{TRES} spectra and radial velocities derived using \texttt{TRICOR}, and involved radial velocity fitting, flux-ratio analysis, spectral disentanglement using both \texttt{CRES} and \texttt{FDBinary}, and atmospheric modelling and spectral synthesis of the disentangled spectrum of Aa using \gssp. The radial velocity fitting and spectral disentangling recover orbital properties of Aa, Ba, and Bb that are in excellent agreement; with the exception of $a \sin i$ values, these orbital properties are also in agreement with previous analyses \citep{heliminiak2017}.
Our \gssp\ modelling of the disentangled spectrum of Aa provides constraints on the effective temperature ($T_{\rm eff} = 7280\pm70$~K), surface gravity (log$(g) = 4.14\pm 0.18$~dex), micro-turbulent velocity ($v_{\rm micro} = 3.61\pm0.19~\kms$), rotation velocity ($v \sin i = 127 \pm 4$~\kms), and metallicity ({[}M/H{]} $ = -0.23 \pm 0.06$). These values are are consistent with \cite{niemczura2015}.

We spent significant effort in attempting to explore the effect of the light fraction of Aa on our results. All aspects of our spectroscopic analysis point to a light fraction between 0.92 and 0.94, while our eclipse modelling favours a lower light fraction more consistent with \citep{heliminiak2017}'s determination of 0.85. We find that our spectroscopic results are quite sensitive to the light fraction, while an excellent fit to the light curve is able to be obtained from our eclipse modelling regardless of whether low (0.83-0.87) or high (0.92-0.96) light fraction bounds are enforced. Therefore, we conclude that our spectroscopic light fraction is the more meaningful value.

The stellar properties of Aa we arrive at are generally not in agreement with the properties implied by \cite{heliminiak2017}'s isochrone fits; given the high degree of modelling uncertainty inherent to isochrone fitting, it is unclear whether any conclusion can be drawn from this discrepancy regarding the co-evolution state of KIC~4150611 as a heptuple.

\begin{acknowledgements}
In addition to the invaluable contributions of the co-authors, the authors wish to acknowledge previous unpublished work by Josh Carter on this system, and thank Nicholas Jannsen, Vincent Vanlaer, and Dominic Bowman for their useful discussion and input. The authors also wish to thank the anonymous referee for their constructive feedback.
The research leading to these results has received funding from the KU Leuven Research Council (grant C16/18/005: PARADISE), and from the Research Foundation Flanders (FWO) under grant agreements G089422N (AT) and 1124321N (LIJ).
CA acknowledges funding by the European Research Council under grant ERC SyG 101071505. Funded by the European Union. Views and opinions expressed are however those of the author(s) only and do not necessarily reflect those of the European Union or the European Research Council. Neither the European Union nor the granting authority can be held responsible for them. This paper includes data collected by the Kepler mission, which are publicly available from the Mikulski Archive for Space Telescopes (MAST) at the Space Telescope Science Institute (STScI). Funding for the Kepler mission is provided by the NASA Science Mission Directorate. STScI is operated by the Association of Universities for Research in Astronomy, Inc., under NASA contract NAS 5–26555.

\end{acknowledgements}

%
%
\bibliographystyle{aa}
\bibliography{bibfile.bib}

\appendix

\section{Eclipse modelling comparison to \textit{TESS} and short-cadence \kepler\ photometry}

\begin{figure*}
   \centering
   \includegraphics[scale=0.7]{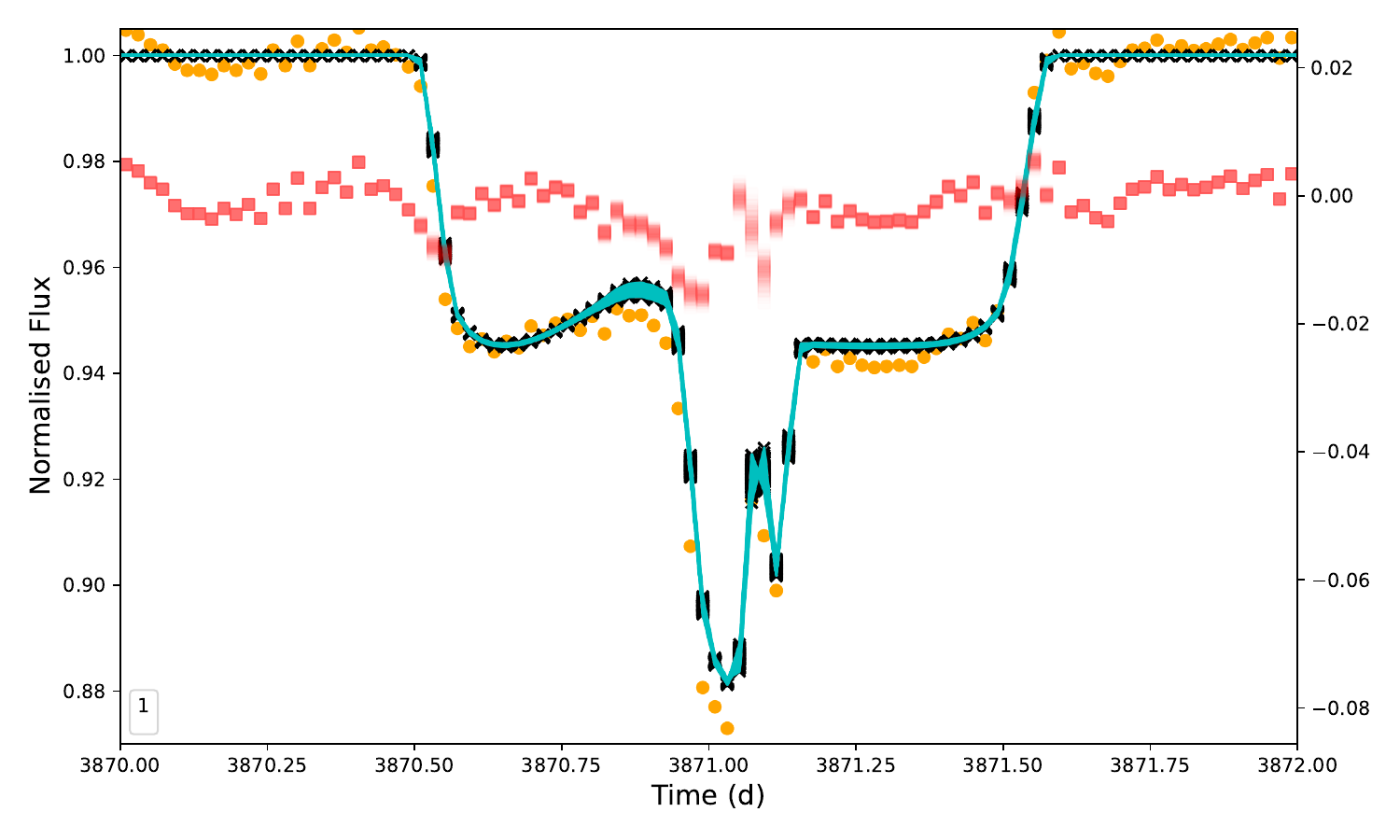}
   \caption{100 randomly selected models from the converged stage of our MCMC simulation for 0.92-0.96 \lfrac. Orange is observational data from the QLP pipeline from \textit{TESS}, black/cyan is the model predictions, and red are the residuals at each timestamp. The primary y-axes are the normalised flux, with the secondary being the residuals.}
   \label{fig:esc9296tess}
\end{figure*}

\begin{figure*}
   \centering
   \includegraphics[scale=0.33]{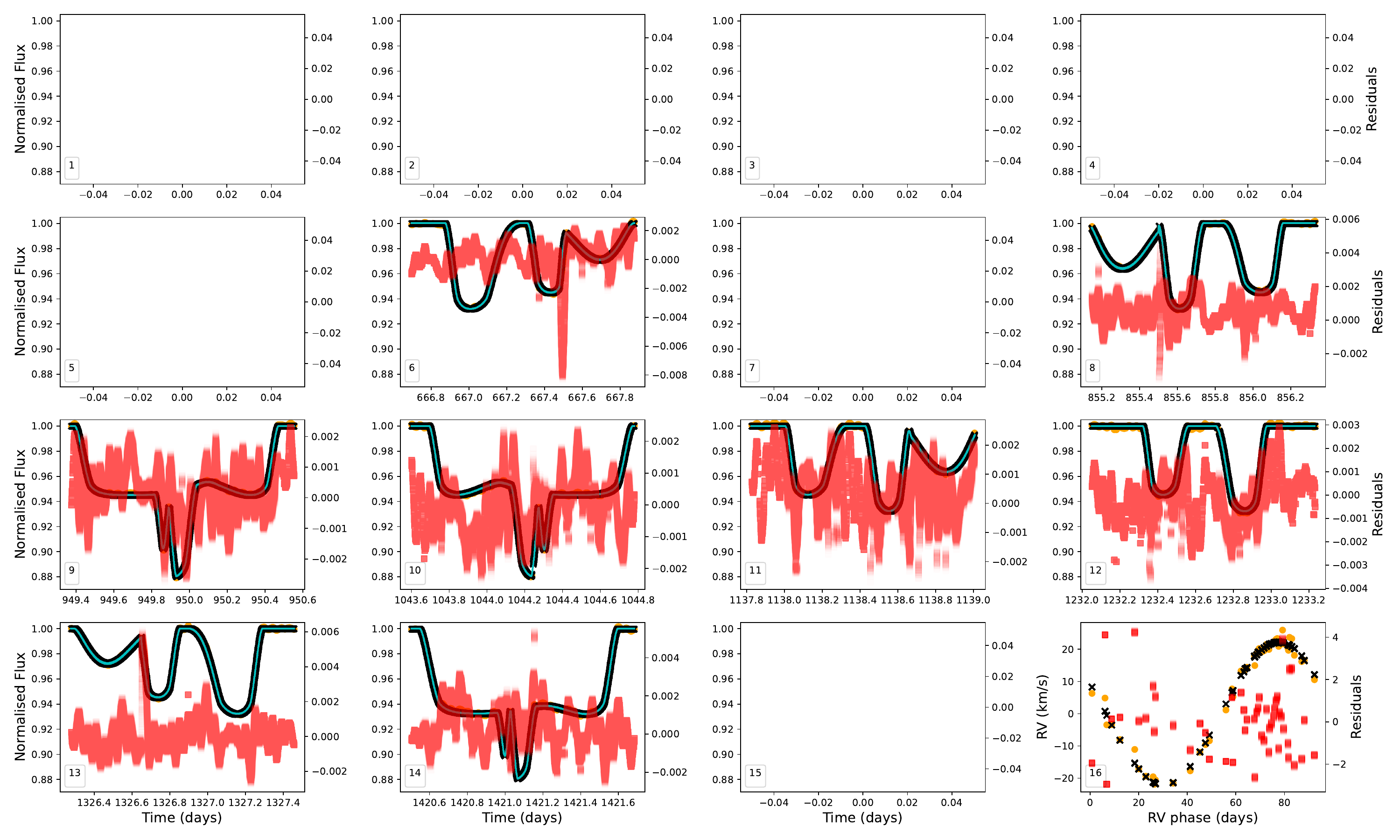}
   \caption{100 randomly selected models from the converged stage of our MCMC simulation for 0.92-0.96 \lfrac. Orange is short-cadence \kepler\ data, black/cyan is the model predictions, and red are the residuals at each timestamp. The primary y-axes are the normalised flux, with the secondary being the residuals.}
   \label{fig:esc9296sc}
\end{figure*}

\section{GSSP fits to disentangled spectra}

\begin{figure*}[h]
   \centering
   \includegraphics[scale=0.9]{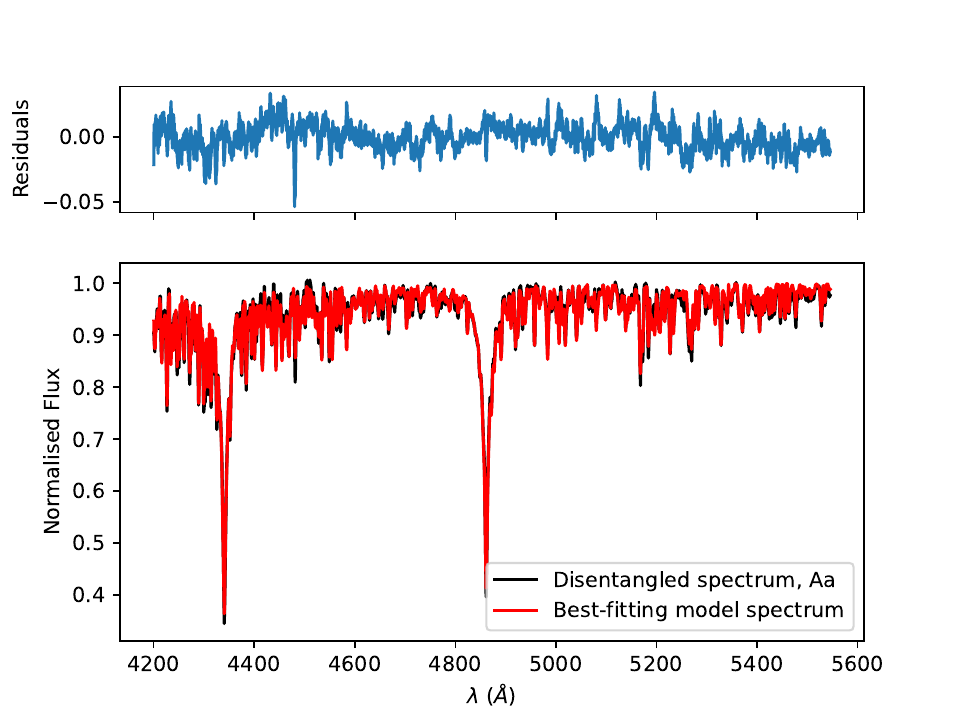}
   \caption{\gssp\ fit to the disentangled spectrum of Aa spectra. Stellar parameters and uncertainties are shown in Table \ref{tab:GSSPfitAa1b}. }
   \label{fig:GSSPfitAa}
\end{figure*}

\begin{figure*}[h]
   \centering
   \includegraphics[scale=0.9]{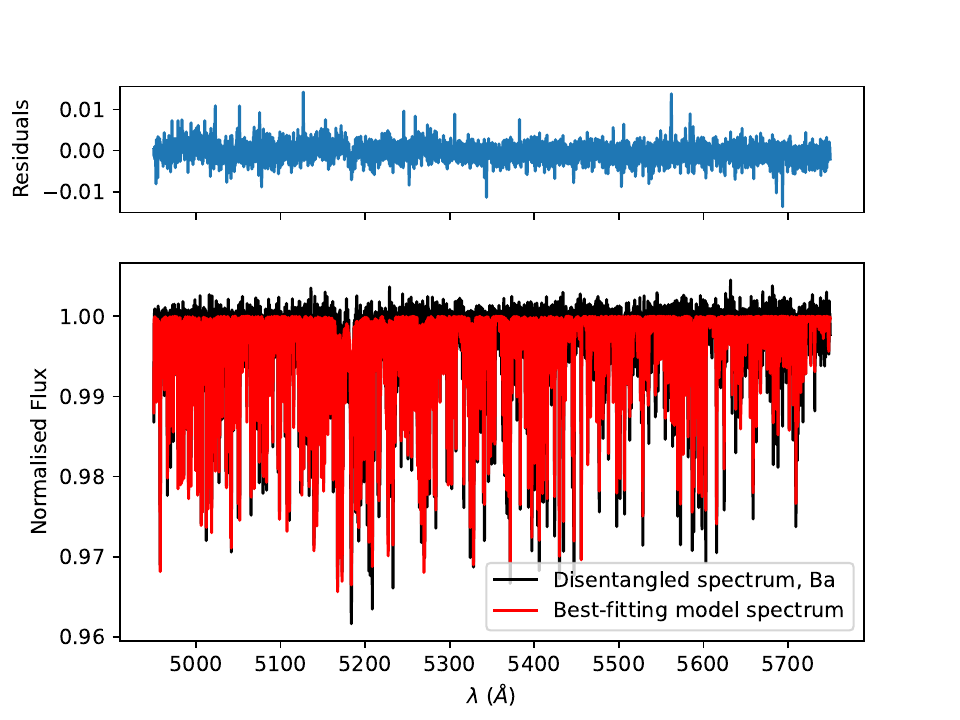}
   \caption{\gssp\ fit to the disentangled spectrum of the brighter of the G stars in the G binary (Ba). Stellar parameters and uncertainties are shown in Table \ref{tab:GSSPfitBa}.}
   \label{fig:GSSPfitBa}
\end{figure*}

\begin{figure*}[h]
   \centering
   \includegraphics[scale=0.9]{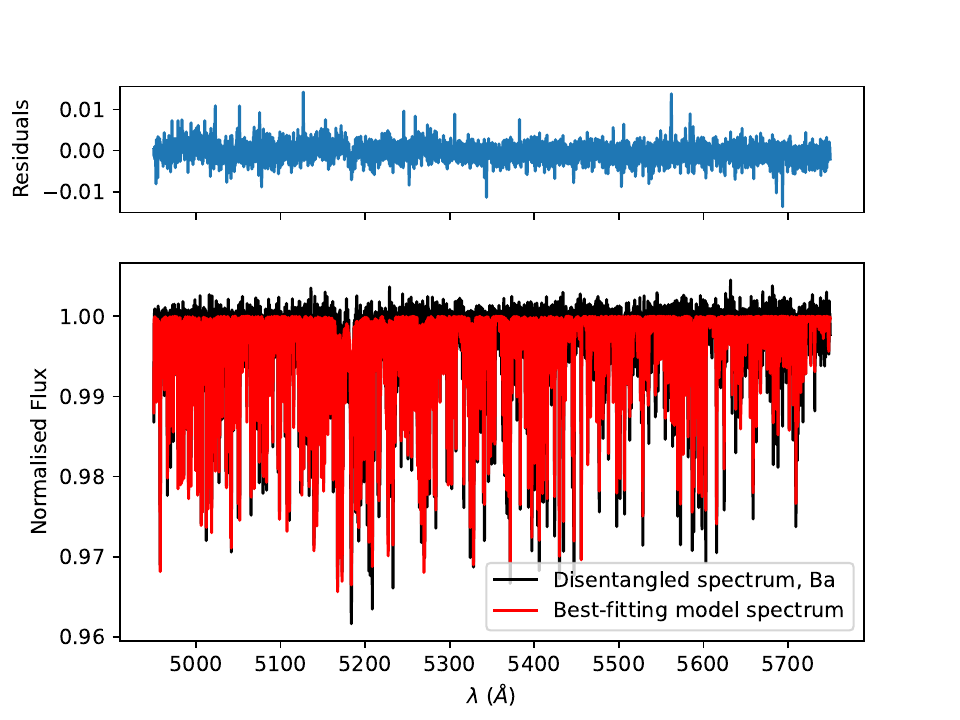}
   \caption{\gssp\ fit to the disentangled spectrum of the dimmer of the G stars in the B binary (Bb). Stellar parameters and uncertainties are shown in Table \ref{tab:GSSPfitBb}.}
   \label{fig:GSSPfitBb}
\end{figure*}

\section{Radial Velocities}

\begin{table*}[h]
\centering
\begin{tabular}{lllllll}
BJD-2,400,000 & RV(Aa) & RV(Ba) & RV(Bb) & errAa & errBa & errBb \\
\hline
55757.7268 & -3.6 & -34.47 & -12.09 & 1.6 & 1.17 & 0.71 \\
55758.7544 & -2.92 & -59.84 & 13.28 & 1.45 & 1.14 & 0.7 \\
55759.6962 & -3.66 & -68.19 & 21.65 & 2.17 & 1.97 & 1.21 \\
55760.9398 & -0.32 & -59.83 & 16.08 & 1.32 & 2.79 & 1.71 \\
55763.7936 & -2.55 & 25.31 & -71.79 & 1.65 & 0.9 & 0.55 \\
55764.7272 & -2.96 & 69.92 & -115.87 & 2.66 & 0.95 & 0.58 \\
55765.7991 & -8.13 & -5.32 & -43.86 & 1.92 & 2.22 & 1.36 \\
55768.8199 & -10.08 & -66.75 & 19.71 & 1.53 & 0.87 & 0.54 \\
55840.5911 & -11.84 & -27.02 & -19.07 & 1.37 & 1.44 & 0.88 \\
55843.7807 & -11.32 & -9.57 & -37.78 & 1.55 & 2.17 & 1.33 \\
55844.6017 & -7.52 & -46.33 & -0.38 & 1.6 & 1.59 & 0.97 \\
55845.6044 & -6.03 & -64.7 & 16.05 & 1.4 & 1.6 & 0.98 \\
55846.6298 & -7.11 & -66.22 & 20.28 & 1.23 & 1.49 & 0.91 \\
55847.6042 & -6.73 & -59.29 & 13.63 & 1.52 & 1.9 & 1.17 \\
55848.5886 & -4.73 & -45.75 & -2.07 & 1.29 & 1.53 & 0.94 \\
55849.6232 & -6.31 & -12.2 & -32.95 & 1.76 & 1.55 & 0.95 \\
55850.61 & -4.62 & 43.79 & -91.55 & 1.38 & 1.47 & 0.9 \\
55851.6013 & -4.03 & 56.32 & -101.52 & 2.06 & 1.55 & 0.95 \\
55852.5932 & -3.15 & -19.51 & -28.7 & 1.4 & 1.34 & 0.82 \\
55853.627 & -5.36 & -55.31 & 8.73 & 1.5 & 1.83 & 1.12 \\
55854.6489 & -3.89 & -65.18 & 19.45 & 1.42 & 1.47 & 0.9 \\
55855.6449 & -4.62 & -63.45 & 18.81 & 1.59 & 1.56 & 0.95 \\
55857.6544 & -6.58 & -34.93 & -11.93 & 1.81 & 1.8 & 1.1 \\
55888.624 & -37.34 & -60.98 & 15.62 & 2.36 & 2.26 & 1.39 \\
55991.0196 & -48.14 & -15.04 & -28.78 & 1.29 & 0.59 & 0.36 \\
56023.0166 & -18.56 & 3.63 & -51.53 & 1.69 & 1.06 & 0.65 \\
56027.9082 & -12.71 & -68.44 & 19.8 & 1.99 & 2.14 & 1.31 \\
56056.9654 & -15.71 & -22.88 & -23.01 & 1.35 & 0.82 & 0.51 \\
56078.8927 & -43.36 & -59.31 & 13.55 & 1.41 & 1.03 & 0.63 \\
56081.8227 & -45.64 & -51.46 & 6.27 & 1.25 & 0.86 & 0.53 \\
56084.7996 & -45.81 & 67.79 & -115.55 & 1.67 & 0.78 & 0.47 \\
56114.8087 & -25.1 & -66.45 & 20.21 & 1.64 & 0.86 & 0.53 \\
56117.7147 & -20.97 & -18.66 & -27.46 & 1.36 & 0.9 & 0.55 \\
56132.7468 & -5.01 & -64.29 & 17.54 & 1.45 & 0.86 & 0.53 \\
56200.6798 & -35.83 & -64.89 & 19.45 & 1.56 & 0.89 & 0.55 \\
56234.6703 & -5.95 & -59.61 & 13.25 & 1.57 & 1.02 & 0.63 \\
56253.5865 & -21.39 & -66.13 & 18.9 & 1.5 & 0.81 & 0.49 \\
56354.0248 & -34.27 & 12.16 & -58.49 & 1.29 & 0.87 & 0.53 \\
56375.9539 & -47.81 & -53.43 & 7.45 & 1.48 & 0.9 & 0.55 \\
56382.9907 & -43.97 & -65.83 & 19.9 & 1.54 & 0.73 & 0.45 \\
56386.9837 & -38.2 & 35.39 & -83.52 & 1.17 & 0.9 & 0.55 \\
56390.8936 & -34.65 & -65.31 & 17.37 & 1.58 & 0.85 & 0.52 \\
56403.9158 & -12.99 & 9.3 & -57.63 & 1.43 & 0.97 & 0.59 \\
56410.9599 & -6.53 & -46.18 & 0.07 & 1.38 & 0.71 & 0.44 \\
56429.972 & -9.69 & 18.38 & -64.54 & 1.57 & 0.76 & 0.47 \\
56436.8268 & -19.95 & -47.83 & 2.33 & 1.5 & 0.77 & 0.47 \\
56442.895 & -29.71 & -64.92 & 18.38 & 1.24 & 0.85 & 0.52 \\
56444.9372 & -29.63 & -57.03 & 10.96 & 1.23 & 0.76 & 0.46 \\
56462.9433 & -46.76 & -45.01 & -1.71 & 1.64 & 0.88 & 0.54 \\
56499.8601 & -12.71 & 58.88 & -105.9 & 1.58 & 1.25 & 0.77 \\
56503.704 & -8.26 & -66.34 & 20.71 & 1.57 & 0.9 & 0.55
\end{tabular}
\caption{Radial velocities obtained from the \textit{TRES} spectra for Aa, Ba, and Bb; units are in \kms.}
\label{tab:rvs_torres}
\end{table*}

\end{document}